\documentclass[twocolumn]{aastex61}

\usepackage{amsmath}

\newcommand\aastex{AAS\TeX}

\shorttitle{\aastex\ Radii of Rapidly Rotating M Dwarfs}
\shortauthors{Kesseli et al.}

\begin{document}

\title{Magnetic Inflation and Stellar Mass II: On the Radii of Single, Rapidly Rotating, Fully Convective M Dwarf Stars}

\correspondingauthor{Aurora Y. Kesseli}
\email{aurorak@bu.edu}

\author[0000-0002-3239-5989]{Aurora Y. Kesseli}
\altaffiliation{Visiting Astronomer at the Infrared Telescope Facility, which is operated by the University of Hawaii under contract NNH14CK55B with the National Aeronautics and Space Administration.}
\affiliation{Department of Astronomy \& Institute for Astrophysical Research, Boston University, 725 Commonwealth Ave., Boston, MA 02215, USA}

\author[0000-0002-0638-8822]{Philip S. Muirhead}
\affiliation{Department of Astronomy \& Institute for Astrophysical Research, Boston University, 725 Commonwealth Ave., Boston, MA 02215, USA}

\author[0000-0003-3654-1602]{Andrew W Mann}
\altaffiliation{Hubble Fellow}
\affiliation{Department of Astronomy, Columbia University, 550 West 120th St., New York, New York 10027, USA}
\affiliation{Department of Astronomy, The University of Texas at Austin, 2515 Speedway, Stop C1400, Austin, TX 78712, USA}

\author[0000-0001-7875-6391]{Greg Mace}
\affiliation{Department of Astronomy, The University of Texas at Austin, 2515 Speedway, Stop C1400, Austin, TX 78712, USA}

\begin{abstract}

Main sequence, fully-convective M dwarfs in eclipsing binaries are observed to be larger than stellar evolutionary models predict by as much as $10-15\%$. A proposed explanation for this discrepancy involves effects from strong magnetic fields, induced by rapid-rotation via the dynamo process. Although, a handful of single, slowly-rotating M dwarfs with radius measurements from interferometry also appear to be larger than models predict, suggesting that rotation or binarity specifically may not be the sole cause of the discrepancy.  We test whether single, rapidly rotating, fully convective stars are also larger than expected by measuring their $R \sin i$ distribution.  We combine photometric rotation periods from the literature with rotational broadening ($v \sin i$) measurements reported in this work for a sample of 88 rapidly rotating M dwarf stars. Using a Bayesian framework, we find that stellar evolutionary models underestimate the radii by $10-15\% \substack{+3 \\ -2.5}$, but that at higher masses ($0.18<M<0.4 M_{Sun}$) the discrepancy is only about 6\% and comparable to results from interferometry and eclipsing binaries. At the lowest masses ($0.08<M<0.18 M_{Sun}$), we find the discrepancy between observations and theory is $13-18\%$, and we argue that the discrepancy is unlikely to be due to effects from age.  Furthermore, we find no statistically significant radius discrepancy between our sample and the handful of M dwarfs with interferometric radii.  We conclude that neither rotation nor binarity is responsible for the inflated radii of fully convective M dwarfs, and that all fully-convective M dwarfs are larger than models predict.

\end{abstract}

\section{Introduction}

M dwarf stars are the most abundant stars in the Galaxy, comprising over 70\% of all stars by number \citep{bochanski10}, yet their fundamental parameters are not well constrained. Radii are particularly difficult to determine because M dwarf stars are intrinsically small and faint, leading to only a few direct radius measurements using long-baseline interferometry \citep[$< 20$, and only two with spectral types later than M3.5; ][]{segransan03, demory09, boyajian12, vonBraun14}. Other M dwarf radius measurements come from eclipsing binary stars (EBs). However, many of these systems reveal radii that are as much as $10 - 15$\% larger than theoretical predictions from stellar evolutionary models, and are on average inflated by $\sim5\%$ \citep[e.g., ][]{torres02, kraus11, han17}.

The inflated radii of M dwarf stars present a problem for exoplanet characterization. The radius precision of a transiting exoplanet is limited by the precision of the stellar radius. The transition from Earth-like planets to Neptune-like planets is believed to occur around $1.5R_{E}$ \citep{rogers15}. If stellar radii are in error by up to 15\%, based on simulations of planet occurrence rates expected for \textit{TESS} \citep{sullivan15}, a significant fraction of the future super-Earth sized planets expected to be discovered by \textit{TESS} would in fact be mini-Neptunes. The errors are even more important when determining planet densities; a 10\% adjustment to the radius of any transiting extrasolar planet results in a 30\% adjustment to the measured exoplanet average density. A 30\% difference in inferred average density is the difference between a rocky or metal dominated interior and would dramatically change the mass fraction attributed to a gaseous envelope. 
Radii of M dwarf stars will be particularly important for \textit{TESS}.  With a 30 day baseline for photometric observations for most of the sky, the majority of the discovered exoplanets in the habitable zone will be around M dwarf stars \citep{muirhead17}.

Several studies have proposed that the larger-than-expected radii of M dwarf stars in EBs are a result of activity and enhanced magnetic fields \citep[often around a few kiloGauss for M dwarf stars;][]{donati06, lopez07, chabrier07}. 
Magnetic field strength and magnetic activity have long been known to be coupled to rotation \citep{parker55}, and more recent observations affirm that M dwarf stars with rotation periods less than $\sim$5 days all show evidence of magnetic activity through chromospheric emission \citep[e.g., ][]{west15, newton17}. In this scenario, EBs are preferentially inflated because of observational biases: they tend to have short orbital periods ($P<5$ days) and are correspondingly synchronously rotating. To account for inflation suggested by this theory, studies such as \citet{kraus11} have suggested adding a rotation parameter into Mass$-$Radius relations for M dwarf stars. 

A large fraction of \textit{single} M dwarf stars are also found to be rapid rotators.  \citet{newton16} found that more than one third of the mid-to-late M dwarf stars in the MEarth survey have rotational periods less than one day.  
If rotation-induced magnetic fields cause larger-than-expected radii in EBs, then a large number of single stars should also have larger-than-expected radii. As of now we do not have a sample of rapidly rotating single stars with precise radius measurements; the mid-to-late M dwarf stars for which interferometric radii measurements are available \citep[Proxima Centauri and Barnard's Star;][]{boyajian12} have rotation periods around 80-130 days \citep{benedict98}.

Alternatively, the inflation may solely be an effect present in EBs.  Disk disruption and/or tidal effects from close binaries could alter the evolutionary history of EBs such that rotation is not the key factor responsible for the larger-than-expected radii \citep{meibom06, morgan12}. \citet{morales10} showed that if magnetic cool spots on active M dwarf stars are preferentially distributed near the poles \citep[as seen by ][]{granzer00, jeffers07, stass09}, the radii could be overestimated by up to 6\% by parameter extraction codes that assume circular stellar disks when modeling EB light curves. Also, reanalysis of EB data from multiple groups has oftentimes lead to vastly different stellar parameters, calling into question the accuracy of parameters extracted from EBs \citep{han17}.

There are a range of possible mechanisms for how ignoring magnetic fields in the models leads to underestimated radii. \citet{chabrier07} used stellar modeling code to demonstrate that rotation-induced surface magnetic fields can lead to larger radii of low-mass stars by two scenarios: (1) strong magnetic fields inhibit convective flows (modeled by decreasing the mixing length parameter), and (2) large magnetic cool spots decrease the overall effective temperature of the star, and thus increase the radius since the luminosity is unchanged.
\citet{chabrier07} predict that only stars above the fully convective boundary would be affected by scenario (1), because their interiors are nearly adiabatic and decreasing the mixing length parameter has little effect. \citet{chabrier07} also showed that scenario (2) alone could inflate the radii of M dwarf stars seen in EBs, but only with a large spot covering fraction of 30-50\% of the stellar surface. 

\citet{feiden14} used the Dartmouth Magnetic Stellar Evolution Tracks and Relations \citep[DMSETR; ][]{feiden12} to explore both of these scenarios in more detail. Instead of modeling scenario (1) using a decreased mixing length parameter, they modeled how the magnetic field could stabilize convection, and found that it could inflate the radii of fully convective stars by 5-6\% if extremely strong interior magnetic fields were invoked (40 MG).  However, theoretical predictions of interior field strength concluded that the above-quoted field strengths are unreasonably large \citep{browning16}. On the other hand, \citet{macdonald17} used a similar approach to that of \citet{feiden14}, but found interior magnetic fields strengths on the order of 10 kG could inflate the radii of fully convective stars to a similar degree as seen in EBs. 

To disentangle the roles of the two scenarios proposed to inflate the radii, and to help provide constraints for future modeling, a sample of rapidly rotating, single, fully convective stars needs to be studied to determine the level of inflation present. 

In this paper, we test the role of rapid rotation on M-dwarf radii by measuring the statistical distribution of radii modulated by the inclination ($\sin i$) of 88 single, rapidly rotating M dwarf stars. In Section \ref{stellarSample} we describe the target selection, and in Section \ref{data} we outline our observations and data reduction procedures. Next, we describe how we obtain rotational broadening ($v \sin i$) measurements in Section \ref{vsini}. We explain the Bayesian approach used to determine the mean inflation of our sample in Section \ref{bayesian}, and present the results from the analysis in Section \ref{results}. Lastly, we explore any potential biases that would arise from this method in Section \ref{biases}, and summarize and conclude in Section \ref{conclusions}.

\section{Stellar Sample} 
\label{stellarSample}

To determine the $R \sin i$ distribution, we combined photometric rotation periods of MEarth targets \citep{newton16}, with $v \sin i$ values that we measured in this work. MEarth has been photometrically monitoring close to 2,000 targets selected to be mid-to-late (M3-M6) M dwarf stars since 2008 with a photometric precision of 1.5\% \citep{berta12, dittmann14}.  
Our measurements of $v \sin i$ are obtained through rotational broadening of absorption lines. The broadening is proportional to the rotational velocity ($v$) modulated by the inclination ($\sin i$), and is the dominant source of broadening for rapidly rotating stars. The measured $v \sin i$ and published rotational periods ($P_{\rm rot}$) are related to the stellar radius ($R$) as follows: 

\begin{equation}
   R \sin i = v \sin i \ P_{\rm rot} \ / \ (2 \pi )
\end{equation}

We selected stars that had a secure periodic detection of photometric modulation \citep[class `A' or `B' rotators from ][]{newton16}. We also required the stars to have a period of less than 5 days, to ensure they were all magnetically active and had $v \sin i$ values that we could resolve with our spectrographs. A large portion of the sample have H$\alpha$ measurements, and every star with a measurement is magnetically active \citep{newton17}. 
We only observed stars with $K$-band magnitudes less than 11, since larger magnitudes required significantly longer exposure times and often returned unsatisfactory results. To isolate the single stars, we did not include any stars that were flagged as binaries in \citet{newton17}, which includes both removal of blended or elongated PSFs and sources flagged as being overluminous for their given color. The multiplicity fraction of M dwarf stars is not precisely known, however modern estimates state that $26 \pm 3\%$ of M dwarf stars are multiples \citep{duchene13}, leading us to conclude that binaries and multiples have been removed from our sample. We also visually inspected all the cross-correlation function to look for multiple peaks and only noticed one of our targets was a previously unknown spectroscopic binary (noted in Table \ref{t:obsData}. Finally, because MEarth stars are selected to be mid-to-late M dwarf stars, all of our sample have mass estimates reported in \citet{dittmann14} that put them around or past the fully convective limit ($M_{\star} \lesssim 0.4 M_{\odot}$). After these cuts, we were left with 110 potential targets from \citet{newton16}, 83 of which we observed, and 7 more that had precise $v \sin i$ measurements from the literature (discussed in more detail in Section \ref{litResults}). 

\section{Observations and Data Reduction}
\label{data}

Data were collected between October 2016 and November 2017 using the Immersion GRating INfrared Spectrograph \citep[IGRINS; ][]{park14} on Lowell Observatory's 4.3-meter Discovery Channel Telescope (DCT) at and the 2.7-meter Harlan J. Smith Telescope at McDonald Observatory. We also used iSHELL \citep{rayner16} on NASA's 3.0-meter  Infrared Telescope Facility (IRTF) on Mauna Kea, Hawaii. IGRINS is a high-resolution (R $\simeq$ 45,000) infrared spectrograph that simultaneously collects \textit{H} and \textit{K}-band spectra \citep{mace16}. iSHELL has a spectral resolution of 75,000 at our chosen wavelength region in the \textit{K}-band ($2.26 - 2.55\mu m$). The instrument, telescope and observation date for each target are shown in Table \ref{t:obsData}. Exposure times were estimated in order to achieve a signal-to-noise ratio (SNR) of $\sim$100. We found that spectra with a SNR significantly lower than 100 yielded large uncertainties in our final calculated $v \sin i$ value, and hence less precise radius estimates.

With the spectral resolution of IGRINS and iSHELL, we were able to resolve rotational broadening for $v \sin i$ values larger than $\sim$3$-$4 km s$^{-1}$ and $\sim$1$-$2 km s$^{-1}$, respectively. In order to resolve rotational broadening in the largest number of stars, we used IGRINS to observe stars with rotation periods less than a day ($v_{rot} \gtrsim 10$ km s$^{-1}$) and iSHELL to observe stars with rotation periods between one and five days ($ 3 \lesssim v_{rot} \lesssim 10$ km s$^{-1}$).

We performed the data reduction of IGRINS spectra using the publicly available pipeline \citep{lee17}.\footnote{\url{https://github.com/igrins/plp}} The pipeline automatically performs dark subtraction, flat fielding, and subtracts out sky emission (i.e. OH airglow) using an ABBA nodding pattern. The pipeline also returns a wavelength solution, calculated using the OH emission lines before their removal. The final product is a 1-D spectrum, which is calibrated but still contains telluric absorption features. We completed the data reduction of the iSHELL spectra using the Spextool for iSHELL package\footnote{\url{http://irtfweb.ifa.hawaii.edu/research/dr_resources/}}.  Spextool \citep{cushing04} was originally created for reduction of SpeX data, however has been updated in the newest release to be compatible with iSHELL data. We used the \texttt{xspextool} function to perform the dark subtraction, flat fielding, order tracing and extraction, linearity correction and wavelength extraction. \texttt{xspextool} also returns a wavelength solution calibrated using ThAr lamps.

\begin{figure}[ht]
\begin{center}
\includegraphics[width=\linewidth]{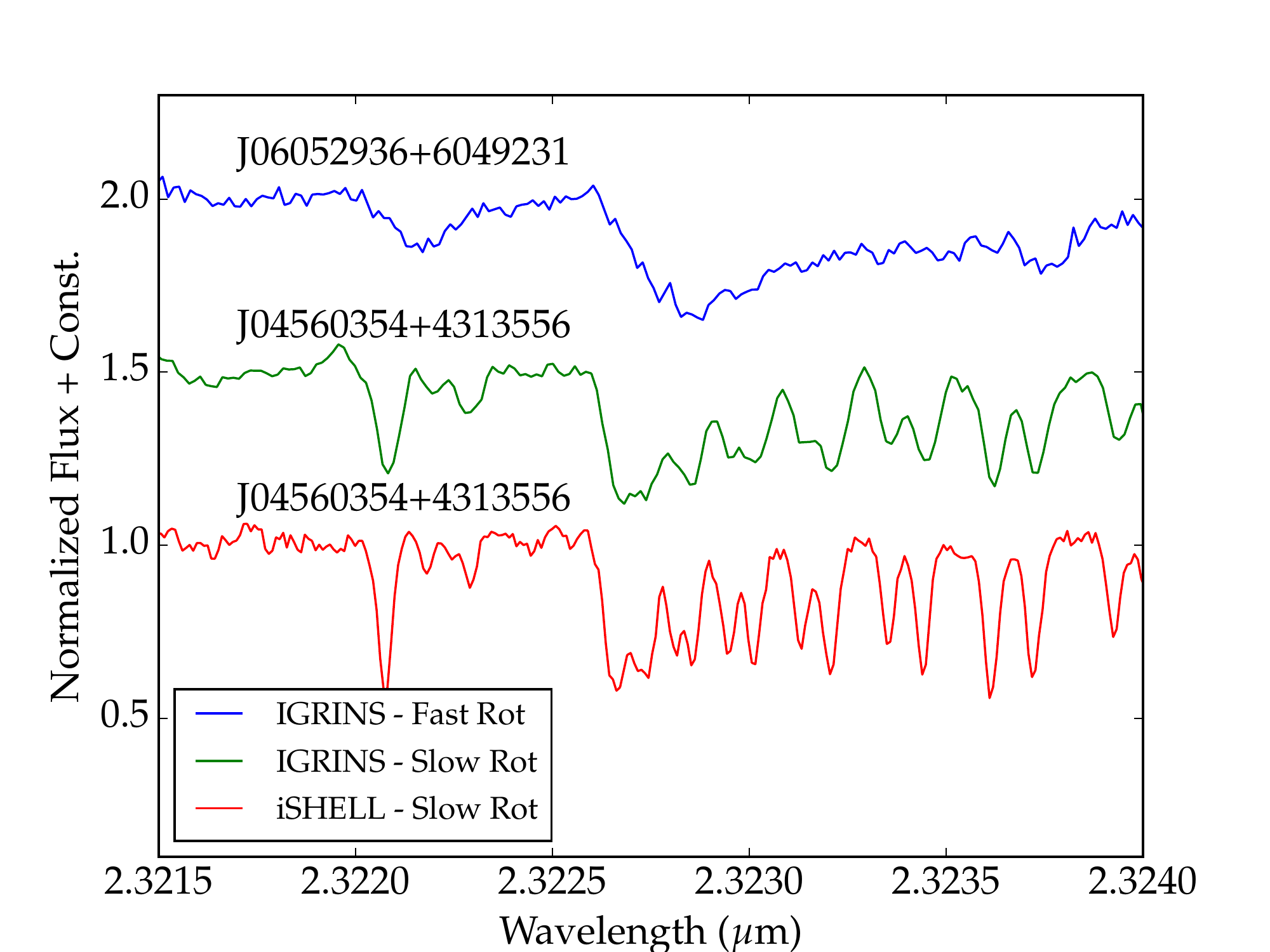}
\caption{\small
IGRINS and iSHELL spectra from our sample, centered on the 1-0 vibrational CO bandhead ($\sim2.3 \mu m$). The blue line shows a rapidly rotating M dwarf with a measured $v \sin i$ of 29.8 km/s (J06052936+6049231) taken with IGRINS. The green line below shows a slowly rotating M dwarf (J04560354+4313556) also taken with IGRINS. The rapidly rotating M dwarf clearly has much broader and shorter absorption lines than the slowly rotating M dwarf. The bottom red line shows the same slowly rotating M dwarf star (again, J04560354+4313556) but taken with iSHELL. The difference in broadening between the two spectra of J04560354+4313556 is entirely due to the resolution difference between the spectrographs. This plot demonstrates why we can observe slower rotators with iSHELL.}
\label{f:spec}
\end{center}
\end{figure}

Large parts of the $H$ and $K$-bands are dominated by telluric lines. We removed telluric absorption features using the \texttt{xtellcor} \citep{vacca03} function, which is also part of the larger Spextool reduction package. Since Spextool is not formatted for IGRINS spectra, we utilized \texttt{xtellcor\_general} for telluric correction of IGRINS spectra. \texttt{xtellcor\_general} can be used with any instrument, given the spectral resolution, an A0 standard spectrum, and a target spectrum. A0 standard stars were taken throughout the night during all observations and were required to deviate in airmass from the target by less than 0.2. 
Examples of our reduced and telluric corrected spectra are shown in Figure \ref{f:spec}.

\section{Determining Rotational Broadening}
\label{vsini}

Our method to determine the $v \sin i$ value is similar to that of many previously published studies \citep[e.g.,][]{west09,muirhead13,reiners17}.
To determine the rotational broadening, we compared the rapidly rotating M dwarf stars to slowly rotating M dwarf stars (P $> 50$ days) also from the \citet{newton16} sample.  In the slowly rotating stars, the rotational broadening is undetectable, and any broadening seen is due to the intrinsic broadening of the spectrograph (see Figure \ref{f:spec} to see how the change in resolution of our two spectrographs broadens the spectra). 
%We perform this analysis using two different methods to ensure our resulting $v \sin i$ values are accurate. In both methods, 

To start, the slow rotators were artificially broadened using the $v \sin i$ kernel, \texttt{rotBroad}, available in the PyAstronomy library.\footnote{\url{https://github.com/sczesla/PyAstronomy}} The rotational broadening kernel requires a linear limb darkening coefficient ($\mu$) as input. We referred to \citet{claret12} to determine the appropriate value of $\mu$, and found that for our sample of stars ($2900 \lesssim T_{eff} \lesssim 3400$) and for $H$ and $K$-band observations, the linear limb darkening coefficient varies between $\sim0.3 - 0.4$. 
So as to not have the choice of limb darkening coefficient bias our final results, we treat it as a nuisance parameter in our Bayesian analysis (see Section \ref{s:nuisance} for details). For all of our reported $v \sin i$ values we use a coefficient of 0.35, since it falls in the middle of the allowed range.

\begin{figure}[ht]
\begin{center}
\includegraphics[width=\linewidth]{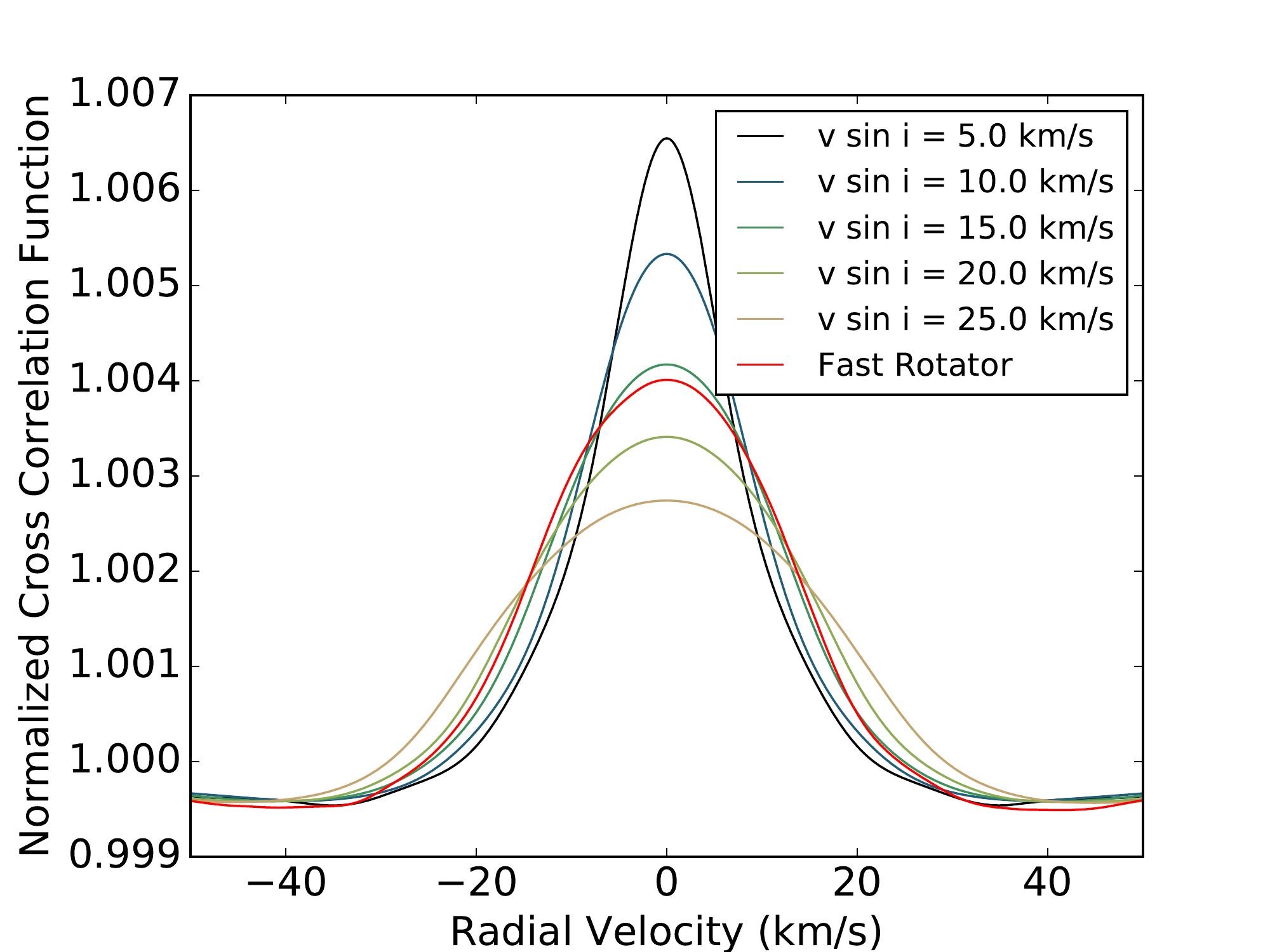}
\includegraphics[width=\linewidth]{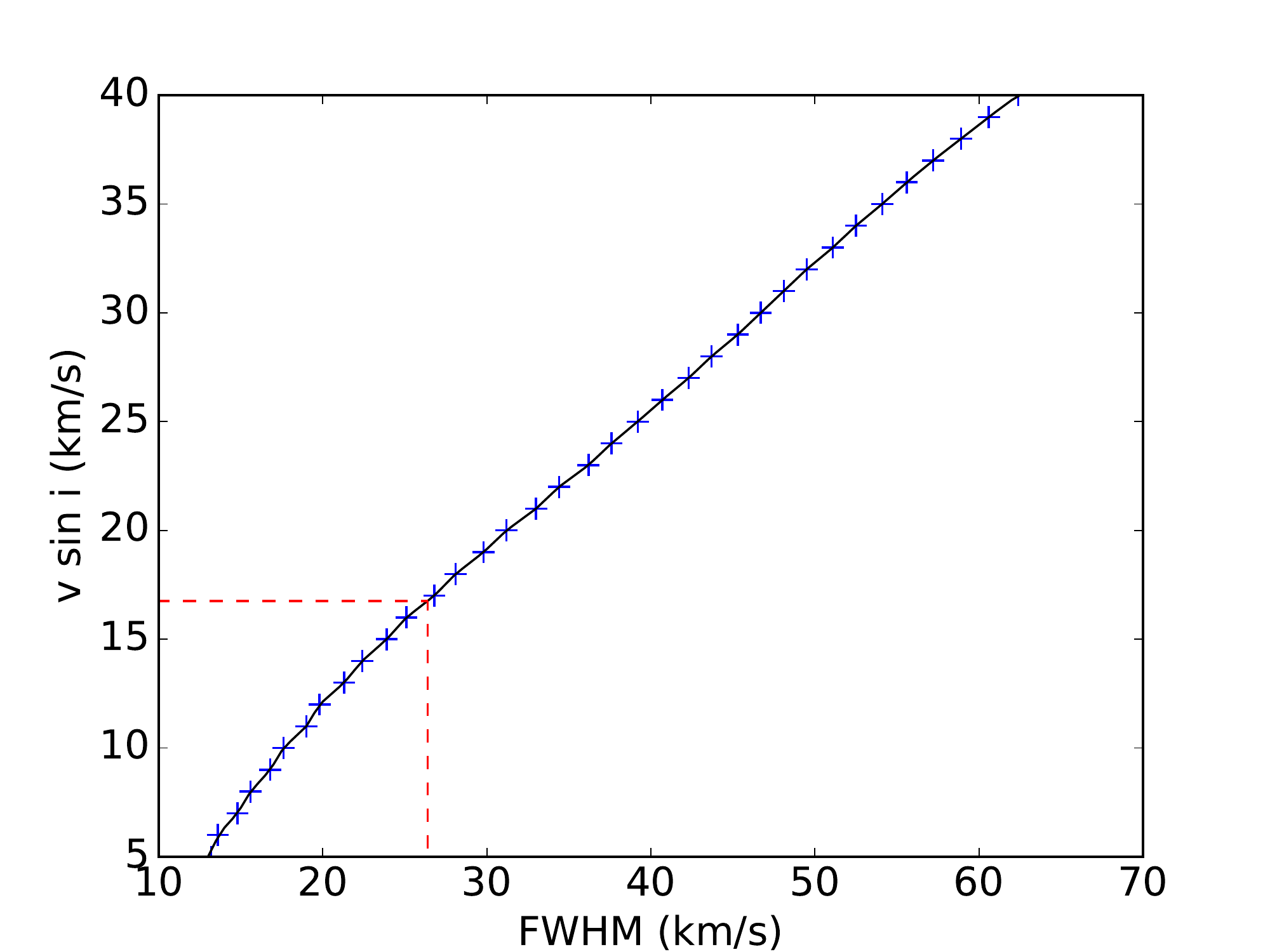}
\caption{\small
\textbf{Top:} Resulting cross-correlation function between a slowly rotating M4.9 \citep{alonso15} dwarf star (J04560354+4313556) and a rapidly rotating M5.9 \citep{shkolnik09} dwarf star (J10204406+0814234), as well as the slow rotator with a few artificially broadened spectra. The darkest blue lines have the smallest $v \sin i$ kernel applied to the slow rotator's spectrum, while the yellow lines used the largest $v \sin i$ kernel. \textbf{Bottom: } Our relation for the measured FWHM versus the $v \sin i$ value for the stars mentioned above. The blue plus signs show the measured FWHM values for the artificially broadened slow rotators, and the black line shows the interpolated relation. The red dashed line shows the measured FWHM of the rapid rotator, and the interpolated $v \sin i$ value. For this specific order we measure a $v \sin i$ of 16.75 km s$^{-1}$. 
}
\label{f:Xcorl}
\end{center}
\end{figure}

 Next, to determine the $v \sin i$ value, the artificially broadened slow rotators were cross-correlated with the original unbroadened spectrum of the slow rotator. The width of the cross-correlation function monotonically increases with increasing rotational broadening. We created a relation between the full width at half maximum (FWHM) of the cross-correlation function to the $v \sin i$ input value of the kernel used to artificially broaden the spectrum. We then cross-correlated the fast rotators to the slowly rotating M dwarf star and interpolated from the FWHM relation to determine a $v \sin i$ value for each fast rotator. Example cross-correlation functions, showing the artificially broadened spectra cross-correlated with the unbroadened spectrum (blue-yellow), and the rapidly rotating target spectrum cross-correlated with the unbroadened spectrum (red), are shown in Figure \ref{f:Xcorl}.

We performed this analysis on individual orders and excluded orders that: 
\begin{itemize}
\item Had low signal-to-noise: the first and last few orders of all spectra are excluded as well as any orders with obvious noise spikes 
\item Were dominated by telluric features 
\item Contained large atomic features (i.e., Na doublet $\sim 2.2 \mu m$), which are subject to non-Gaussian pressure broadening and therefore can lead to over-estimated $v \sin i$ measurements
\end{itemize}

We found that the CO bands ($\sim 2.3 \mu m$) were ideal for this calculation and returned especially precise measurements of $v \sin i$. The relatively high mean molecular weight of CO and the low Land\`e factors for these particular CO transitions reduce the dependence of the line widths on magnetic fields and pressure broadening, respectively. For spectra obtained with iSHELL all of our $v \sin i$ measurements were from orders containing CO band features, and for our IGRINS spectra about half the orders used were dominated by the CO bands. Because of this, we are confident that the broadening we measure is due to rotation and not magnetic or pressure broadening. Uncertainties were calculated from the standard deviation between $v \sin i$ measurements in different orders.

\subsection{Comparison to Previous Results}
\label{litResults}

\begin{figure}
\begin{center}
\includegraphics[width=\linewidth]{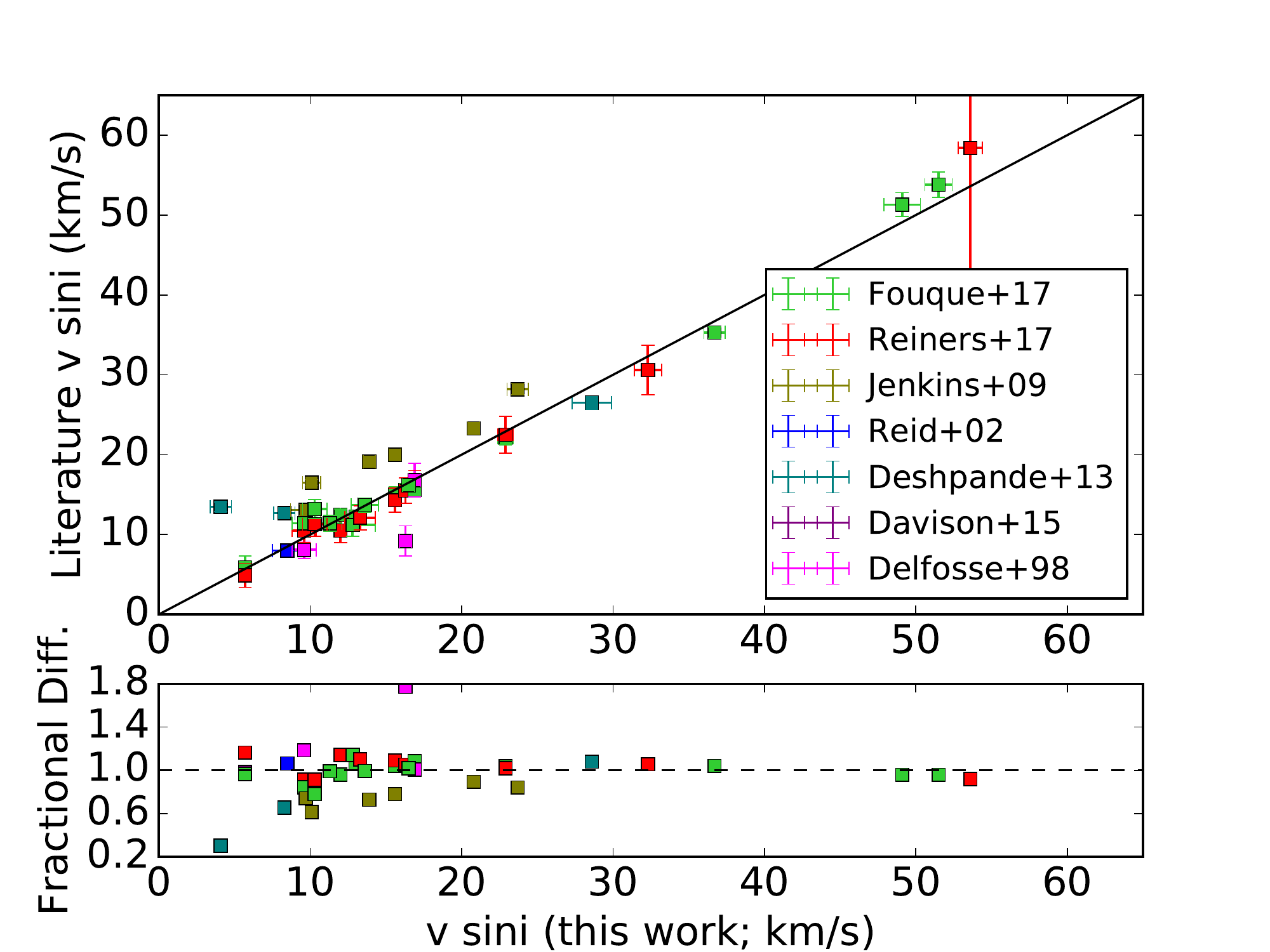}
\caption{\small
Previously recorded literature values of $v \sin i$ compared to $v \sin i$ values measured in this work. The fractional difference in the bottom panel is given by the $v \sin i$ (this work) divided by the $v \sin i$ from the literature. }
\label{f:litCompare}
\end{center}
\end{figure}

Some of our targets have measured $v \sin i$ values in the literature, and we can compare our results to these previous measurements. These results are shown in Figure \ref{f:litCompare} and reported in Table \ref{t:obsData}. Although we found a similar trend in the data, the spread is larger than the reported uncertainties. 
Even with this spread, we are confident in our measurements because we achieved the greatest agreement (74\% of points within 1 sigma, all within 2 sigma) between surveys that use spectrographs with the highest resolution ($R\sim57,000$, \citealt{davison15}; $R>80,000$, \citealt{reiners17}; $R\sim65,000$, \citealt{fouque17}). All the points with greater levels of discrepancy were measurements taken with spectrographs with lower resolution than our survey ($R<45,000$). 

Because of our consistent measurements with both \citet{reiners17} and \citet{fouque17}, we added 7 of their $v \sin i$ measurements to our sample that met all of our criteria listed in Section \ref{stellarSample}, but for which we did not measure a $v \sin i$ value. This increased our total sample to 88 stars. The added targets are listed in Table \ref{t:obsData}.

\section{Bayesian Statistical Analysis}
\label{bayesian}

We combined our measured $v \sin i$ values with rotation periods using equation 1, and in a method similar to that of previous studies \citep{jackson09, jackson16, jackson18}, we determined the average inflation (if any) of the radii of the stars compared to reported radius predictions. Unlike previous studies however we used a completely Bayesian framework for our statistical analysis.

In the following analysis the predicted radius is referred to as $R_{p}$. Table \ref{t:radiiMethods} outlines how we arrived at $R_{p}$ and how we have labeled each method in the following text. All methods began with absolute $K$-band magnitudes ($M_{K}$) for each star, which were determined by combining 2MASS apparent $K$-band magnitudes with parallax measurements reported in \citet{dittmann14}. $M_{K}$ was transformed into a radius directly, or first into a mass (using an $M_{K}$ - Mass relation) and subsequently into a radius (using a Mass-Radius relation). We have also denoted which relations use empirical data and which relations are from stellar evolutionary models. 

The Dartmouth Stellar Evolution Model isochrones utilize the updated 2012 photometric systems and were created using the online Web Tool \footnote{\url{http://stellar.dartmouth.edu/models/isolf_new.html}}. The Padova stellar evolutionary models were obtained using the online\footnote{\url{http://stev.oapd.inaf.it/cgi-bin/cmd_3.0}} for PARSEC v1.0. The Mesa Isochrones and Stellar Tracks (MIST) models were generated using the online web interpolator \footnote{\url{http://waps.cfa.harvard.edu/MIST/interp_isos.html}}. All three of the stellar evolutionary models use a 5 Gyr isochrone and a metallicity of 0.14 dex (the average metallicity of rapid rotators with metallicities estimated in \citet{newton14}). The BHAC model used a 5 Gyr isochrone as well, however super-solar metallicity isochrones are not publicy available so we used the solar metallicity isochrone. We chose a 5 Gyr isochrone because we do not have individual age estimates and previous studies that compared radii to model predictions almost exclusively used this age \citep[e.g., ][]{boyajian12, han17}. We discuss the effect of changing the metallicity and age of the isochrone in Section \ref{DartBias}. 

\begin{deluxetable*}{c c c c c}
\tablecolumns{5}
\centering
\tablewidth{0pt}
\tablecaption{Radius Prediction Methods}%title of the table %title of the table

\tablehead{\colhead{Method Name} & \colhead{$M_{K}$ - Mass} & \colhead{Mass - Radius} & \colhead{$M_{K}$ - Radius} & \colhead{Reported in}
\\ 
\colhead{} & \colhead{Reference} & \colhead{Reference} & \colhead{Reference} & \colhead{}}

\startdata
Benedict+Boyajian & \citet{benedict16}\tablenotemark{*} & \citet{boyajian12}\tablenotemark{*} & None & None \\ 
Mann15 & None & None & \citet{mann15}\tablenotemark{*} & None \\
Dittmann14 & \citet{delfosse00}\tablenotemark{*} & \citet{boyajian12}\tablenotemark{*} & None & \citet{dittmann14}\\ 
Newton16 & \citet{delfosse00}\tablenotemark{*} & \citet{bayless06}\tablenotemark{*} & None & \citet{newton16}\\ 
Benedict+Dartmouth & \citet{benedict16}\tablenotemark{*} & \citet{dotter08}\tablenotemark{$\dagger$} & None & None\\ 
Dartmouth & None & None & \citet{dotter08}\tablenotemark{$\dagger$} & None \\ 
Benedict+Padova & \citet{benedict16}\tablenotemark{*} & \citet{bressan12}\tablenotemark{$\dagger$} & None & None\\ 
Benedict+MIST & \citet{benedict16}\tablenotemark{*} & \citet{choi16}\tablenotemark{$\dagger$} & None & None\\ 
Benedict+BHAC & \citet{benedict16}\tablenotemark{*} & \citet{baraffe15}\tablenotemark{$\dagger$} & None & None\\ 
\enddata

\tablenotetext{*}{Empirically derived relation}
\tablenotetext{$$\dagger$$}{Stellar Evolutionary Model}

\label{t:radiiMethods}
\end{deluxetable*}

Our goal was to estimate by what percentage the radii are inflated given the predicted radii ($R_{p}$)  and the measured $R \sin i$ values of the sample (hereafter, ${\bf R_{p}}$ and ${\bf R {\rm \bf sin} i}$ respectively, in bold to indicate these are arrays of values). We introduce an inflation parameter ($\alpha$), which can take a value ranging from 0.9 to 1.25 (corresponding to a radius inflation of -10\% through 25\%). To simulate different levels of radius inflation, each value in ${\bf R_{p}}$ was multiplied by $\alpha$ then compared to ${\bf R {\rm \bf sin} i}$. 

In Bayesian inference, oftentimes there are parameters in the model that are not parameters of interest--- called nuisance parameters. There are two such parameters in our analysis. The first is a cutoff in the $\sin i$ value. By using spot modulation to determine stellar rotation periods, an inclination bias may have been introduced into the sample since stars with pole-on orientations are not detectable: the spots do not rotate into and out of view and therefore do not cause photometric modulation. We used the variable $\beta$ to represent the cutoff below which we do not measure any $\sin i$ values. The linear limb darkening coefficient ($\mu$), which was used as an input to the $v \sin i$ kernel, was treated as the second nuisance parameter. In Section \ref{s:nuisance} we show how we marginalized over these parameters so they are not included in the final results, but for now we leave them in our analysis.

We determined the most likely value of $\alpha$ using Bayes' theorem. Following the notation of \citet{gregory}, we construct the following form of Bayes' theorem:

\begin{multline}
p(\alpha, \mu, \beta | {\bf R {\rm \bf sin} i}, {\bf R_{p}}) \\ 
= \frac{p(\alpha, \mu, \beta|{\bf R_{p}})\  
p({\bf R {\rm \bf sin} i}|\alpha, \mu, \beta, {\bf R_{p}}) }
{p({\bf R {\rm \bf sin} i}|{\bf R_{p}})} \\
% = \prod \ p(R_{*} \sin i|\alpha, \mu, \beta, R_{p})
= \prod_{j=1}^{88} \ 
\frac{p(\alpha, \mu, \beta|R_{p,j})\  
p((R \sin i)_{j}|\alpha, \mu, \beta, R_{p,j}) }
{p((R \sin i)_{j}|R_{p,j})}
\end{multline} 

\noindent where the subscripted and unbolded symbols represent values for an individual star, which are each multiplied together in the product to get the posterior probability function ($p(\alpha, \mu, \beta | {\bf R {\rm \bf sin} i}, {\bf R_{p}})$). 
The posterior probability function is a probability distribution for different values of $\alpha$ (and $\beta$, $\mu$), given the data (${\bf R {\rm \bf sin} i}$) and assuming that ${\bf R_{p}}$ is correct.
The most likely value of $\alpha$ is given by the peak of the posterior probability function. Any previously known information about inflation can be incorporated into the prior ($p(\alpha|{\bf R_{p}})$). Because we did not have much information on how likely different inflation values were, we used a uniform prior for our analysis (our exact choice of prior is discussed in more detail in Section \ref{results}). The likelihood function is given by $p({\bf R {\rm \bf sin} i}|\alpha, \mu, \beta, {\bf R_{p}})$ and is the probability of obtaining the data; the majority of our effort was in constructing this probability distribution function. Lastly,  $p({\bf R {\rm \bf sin} i}|{\bf R_{p}})$ is the normalization factor and is the integrated probability over all values of $\alpha$, $\mu$ and $\beta$, within their respective prior boundaries.  

\subsection{Constructing the Likelihood Function} 

 To construct the likelihood function, we combined a series of probability distribution functions (PDFs) to determine $ p((R \sin i)_{j}|\alpha, \mu, \beta, R_{p,j})$ for each star. For now we will not discuss the nuisance parameters ($\mu$ and $\beta$), as we cover them in section \ref{s:nuisance}. For the remainder of this section we used the radii obtained using the \citet{dittmann14} method from Table \ref{t:radiiMethods} as $R_{p}$, and in Section \ref{results} we will show the results from other radius predictions.

We followed the formalism of \citet{gregory} for combining PDFs to construct the likelihood functions. We start by defining the variables we used throughout: 

\begin{equation} 
x = \sin i
\end{equation}

\begin{equation} 
y = (R \sin i)_{j}
\end{equation}

\begin{equation} 
z = \alpha \times R_{p,j} \sin i
\end{equation}

\noindent where $x$ and $z$ are variables with different probabilities and $y$ is our measurement.  We calculated the PDFs for $x$ and subsequently $z$. 
The PDFs are written using the notation $f_{X}(x)$, where $X$ is the proposition that the value $x$ is within $x+dx$. 

\begin{equation}
   f_{X} (x) = \frac{x}{\sqrt{1-x^2}} dx
\end{equation}

\noindent $f_{X}(x)$ gives the geometric probability of measuring $\sin i$, assuming a randomly oriented rotational axis with uniform probability over a sphere. The PDF for $z$ is more complicated, and we first combined the PDFs of $R_{p,j}$ and $x$. The PDF of $R_{p,j}$ is a normal Gaussian of the form: 

\begin{equation} 
f_{R_{p}} (r) = \frac{1}{\sqrt{2 \pi \sigma_{r,j}^2}} \exp \frac{- (r - \alpha R_{p,j})^2}{2 \sigma_{r,j}^2}
\end{equation}

\noindent where $\sigma_{r,j}$ is the uncertainty associated with each radius estimate \citep[here 5\%; ][]{dittmann14}. $f_{X}(x)$ and $f_{R_{p}} (r)$ are then combined using a product distribution. 

\begin{equation}
   f_{Z}(z) = \int_{-\infty}^{\infty} f_{X} (x) f_{R_{p}} (z/x) \frac{1}{|x|} dx 
\end{equation}

\noindent We combined the measurement uncertainty associated with $y$ with the PDF to create the final likelihood function for an individual star. We combined our measurement uncertainties with $f_{Z}(z)$ using a convolution given by

\begin{equation}
p((R \sin i)_{j}|\alpha, \mu, \beta, R_{p,j}) = \int_{-\infty}^{\infty} dz f_{Z}(z) f_{E}(y-z)
\end{equation} 

\noindent where $f_{E}(y-z)$ is the PDF of the measurement uncertainty and given by the following normal Gaussian distribution

\begin{equation} 
f_{E}(y-z) = \frac{1}{\sqrt{2 \pi \sigma_{m,j}^2}} \exp \frac{- (y - z)^2}{2 \sigma_{m,j}^2}
\end{equation}
 
\noindent where $\sigma_{m,j}$ is the uncertainty associated with each of our $(R \sin i)_{j}$ measurements, and includes both the propagated uncertainties in our $v \sin i$ measurements and the uncertainties in the periods reported by \citet{newton16}. Plugging Equations 10 and 8 into Equation 9, we obtain the final equation for the likelihood function

\begin{multline}
%\footnotesize
p((R \sin i)_{j}|\alpha, \mu, \beta, R_{p,j}) =\\ \int_{z=-\infty}^{z=\infty}\int_{x=0}^{x=1} \frac{1}{\sqrt{4 \pi^2 \sigma_{m,j}^2\sigma_{r,j}^2}}  \frac{x}{\sqrt{1-x^2}} \frac{1}{|x|}\\
\exp \left( \frac{- (z/x - \alpha R_{p,j})^2 }{2 \sigma_{r,j}^2} + \frac{- (y - z)^2}{ 2 \sigma_{m,j}^2}\right) dz\ dx 
%\end{split} 
\end{multline}

This equation cannot be integrated analytically, so we integrated it numerically using the \texttt{scipy integrate.dblquad} function, which is specifically tailored for numerically integrating double integrals. The value returned by the integral for $p((R \sin i)_{j}|\alpha, \mu, \beta, R_{p,j})$ is the probability of the data given the model for one single $(R \sin i)_{j}$ measurement. We repeated this integration for each object and combined the probabilities by multiplying all the individual probability values together. Then, to construct the likelihood function we again repeated the process for the entire range of $\alpha$ to obtain a probability of measuring the data for each $\alpha$ in the inflation range. 

\subsection{Marginalizing Over Nuisance Parameters} 
\label{s:nuisance}

To remove the nuisance parameters from the final likelihood function, we integrated over them to create a marginalized likelihood function. This is given mathematically by the following:

\begin{multline}
    p(\alpha | {\bf R {\rm \bf sin} i}, {\bf R_{p}})=\\
    \int_{\beta = 0.0}^{\beta=0.4} \int_{\mu=0.3}^{\mu=0.4}  d\beta \ d\mu \ p(\alpha, \mu, \beta | {\bf R {\rm \bf sin} i}, {\bf R_{p}}) 
\end{multline}

\begin{figure}[ht]
\begin{center}
\includegraphics[width=\linewidth]{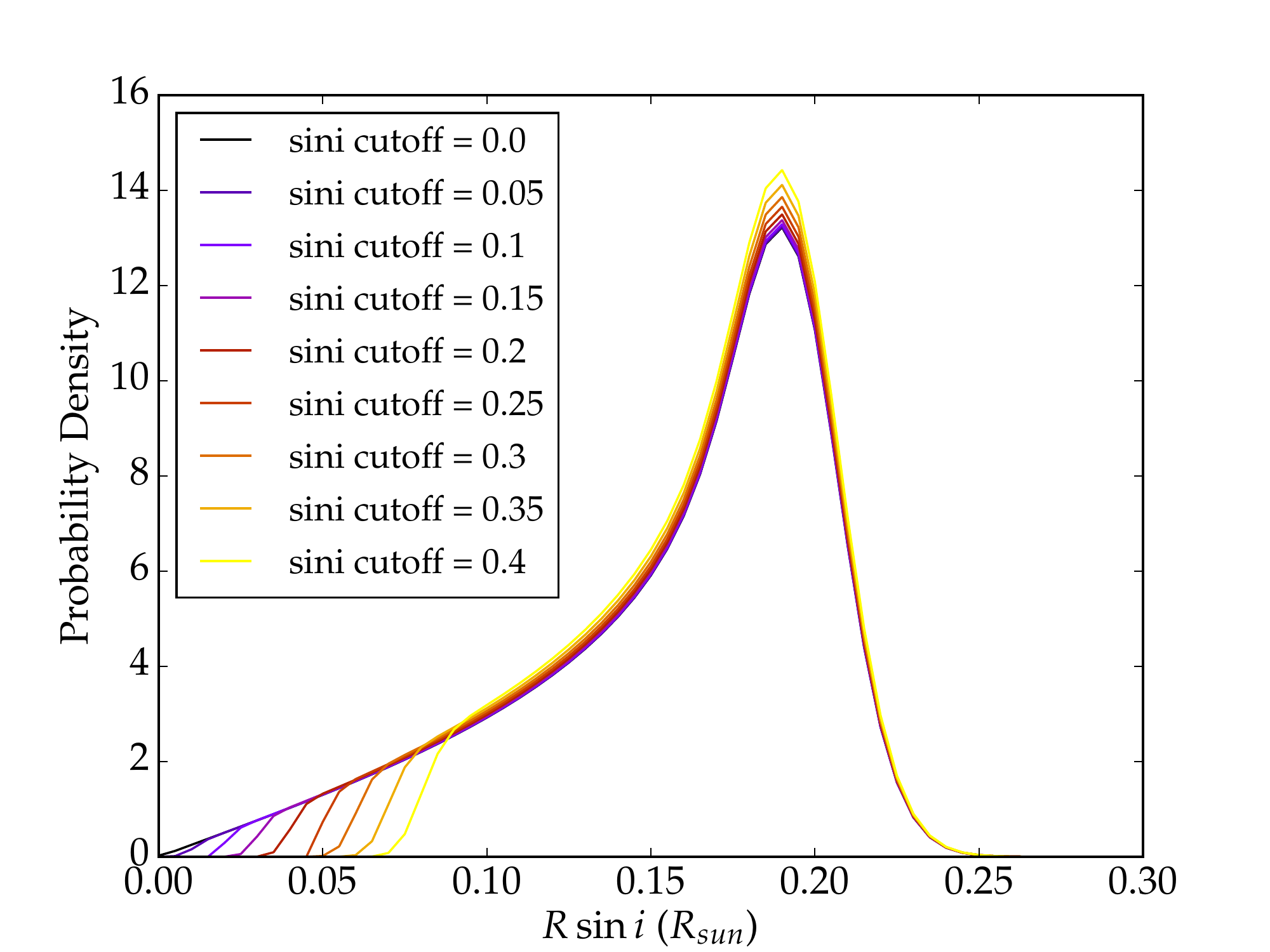}
\caption{\small
Probability distribution functions of $R \sin i$ for a single star. This star was assigned a radius of $0.2R_\odot$. The purple line ($\sin i$ cutoff of 0.0) shows the full $R \sin i$ expected distribution that we used in all our previous analyses. Larger $\sin i$ cutoffs show what the PDF would look like if we assume that the sample from \citet{newton16} did not include stars with inclinations close to pole-on. The larger the $\sin i$ cut off the more biased the sample is against pole-on inclinations.}
\label{f:siniPDFs}
\end{center}
\end{figure}

We first explored the $\sin i$ distribution bias ($\beta$). We cut off the tail of the $\sin i$ PDF used in the likelihood function analysis at a range of $\sin i$ values from 0.0 through 0.4. The PDFs for the $R \sin i$ distributions for a single star are shown in Figure \ref{f:siniPDFs}. 

\begin{figure}[ht]
\begin{center}
\includegraphics[width=\linewidth]{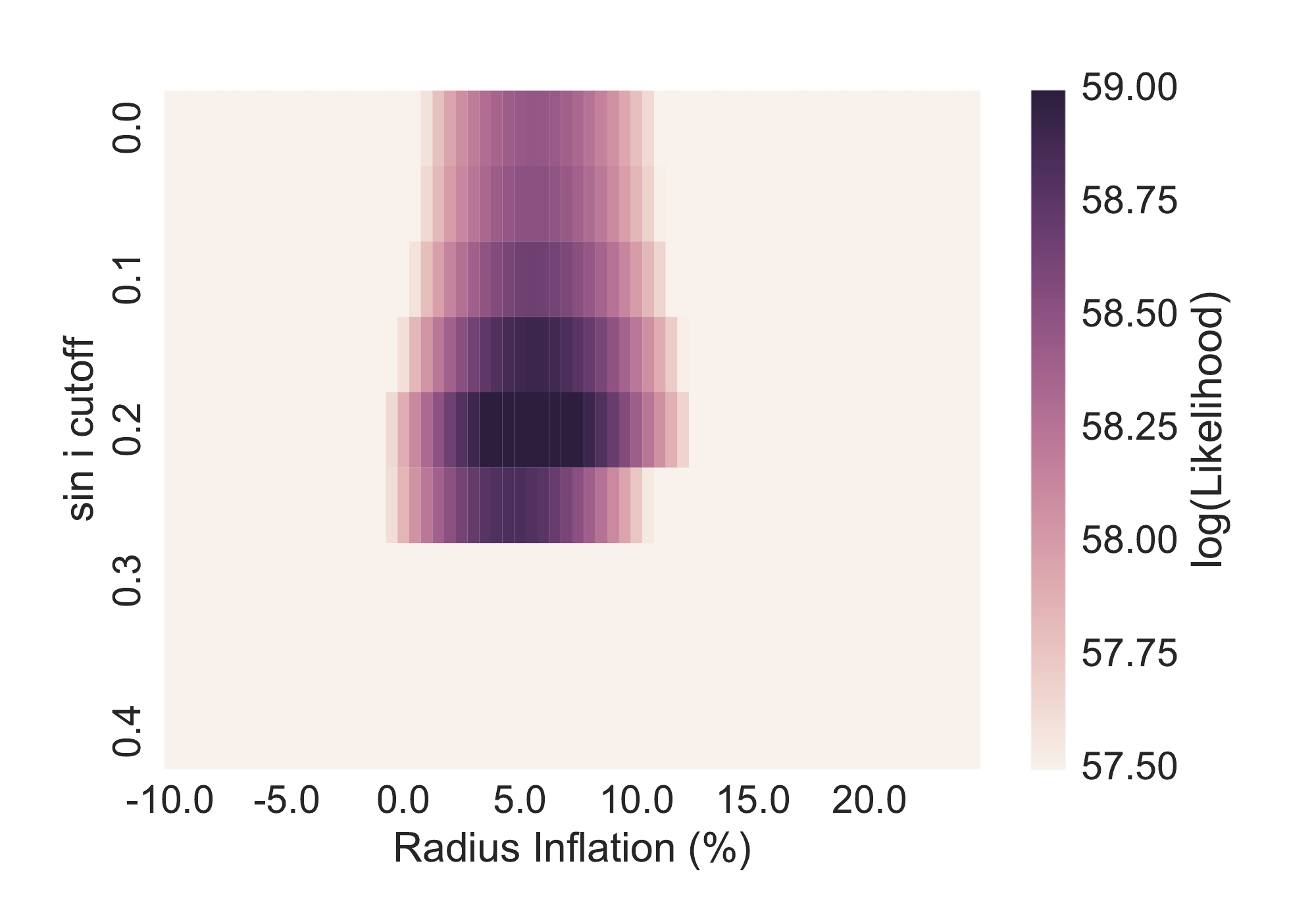}
\caption{\small
Likelihood functions for a range of $\sin i$ cutoff values. By looking at a single $\sin i$ cutoff row, it is clear that the likelihood function peaks around 5\% inflation as we saw before. There is also a maximum probability at a $\sin i$ cutoff of 0.2, with a sharp drop-off after 0.25. }
\label{f:siniLikelihoods}
\end{center}
\end{figure}

Following in our likelihood analysis as before, we created likelihood functions, but this time for a range of $\sin i$ cutoff values. The resulting likelihood functions are shown in Figure \ref{f:siniLikelihoods}. The plot shows that the most likely $\sin i$ cutoff is 0.2, meaning that the \citet{newton16} sample does not include stars with inclinations within $\sim 12 \arcdeg$ of pole-on. Given our data it is unlikely that there exists a $\sin i$ cutoff  $\gtrsim$0.25, and there exists a sharp drop off in probability at this point.  

\begin{figure}[ht]
\begin{center}
\includegraphics[width=\linewidth]{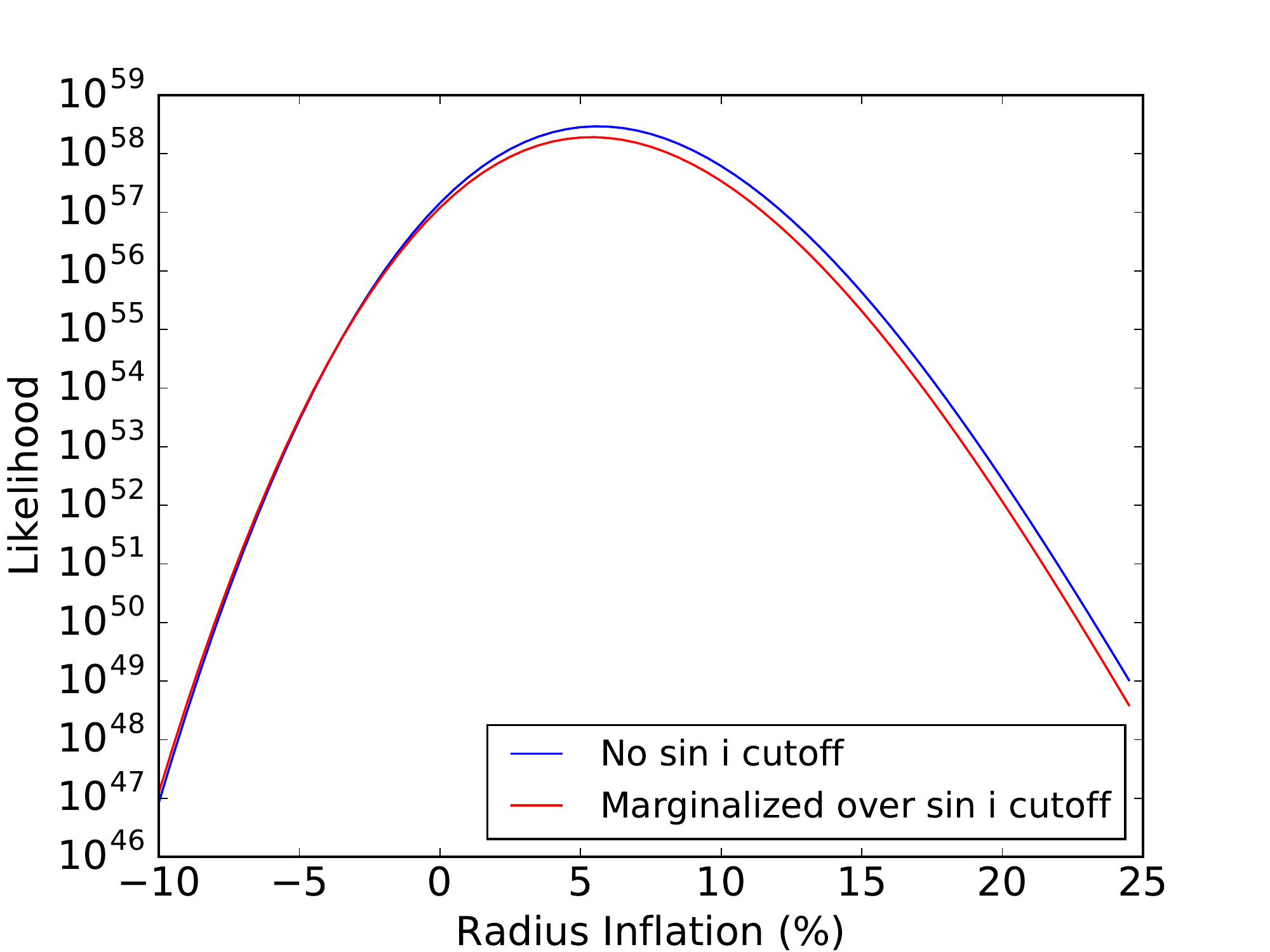}
\caption{\small
 Marginalized likelihood function (red) and original likelihood function (blue). The peak of the likelihood function is shifted slightly to lower radius inflation values for the marginalized likelihood function, however the shift is less than 0.5\% (the resolution of our grid), and both functions peak at the same radius inflation value.}
\label{f:margPDF}
\end{center}
\end{figure}

 We integrated over $\beta$ at each value of $\alpha$ and plot the marginalized likelihood function as well as a likelihood function where we did not consider the effects of a $\sin i$ cutoff value in Figure \ref{f:margPDF}. We found that there seems to be a slight shift in the likelihood function to smaller values of radius inflation, however this shift is smaller than the resolution of our grid and significantly smaller than our error bars.

\begin{figure}[ht]
\begin{center}
\includegraphics[width=\linewidth]{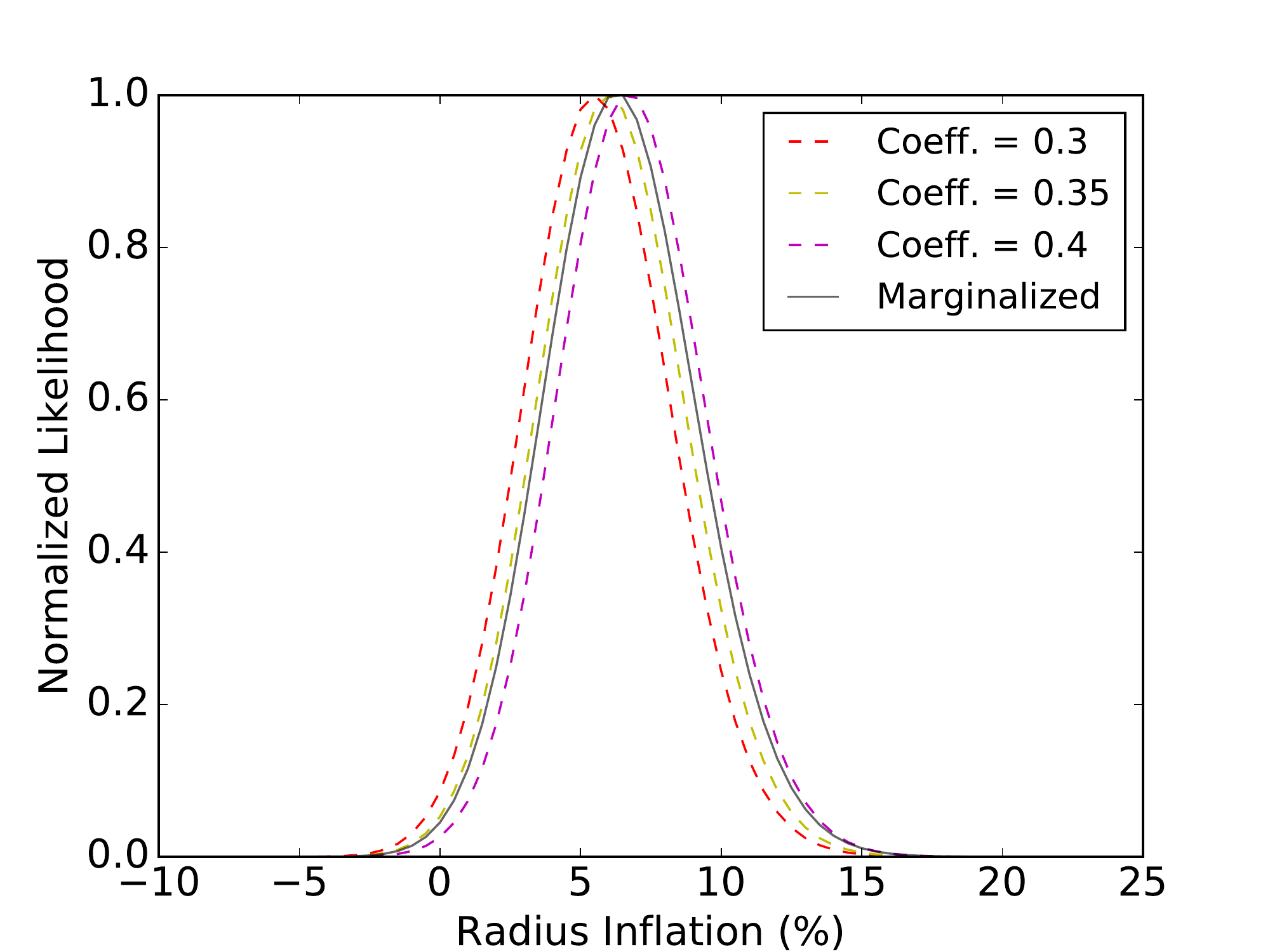}
\caption{\small
 The resulting likelihood functions using a linear limb darkening coefficient of 0.3 (red) and 0.35 (yellow) and 0.4 (magenta), and the likelihood function marginalized over the limb darkening coefficient (grey). By using a limb darkening coefficient at the top and bottom of the range set by our stellar sample, we change the peak likelihood by $\sim 1\%$. This value is within our 1-sigma error bars for the likelihood function.}
\label{f:limbDark}
\end{center}
\end{figure}

We performed the exact same analysis for the limb darkening coefficient as we did for the $\sin i$ cutoff value. 
According to \citet{claret12} our stellar sample covers a range of linear limb darkening coefficients from $\mu \sim 0.3- 0.4$ for observations in $K$-band. We therefore calculated $v \sin i$ values for a range of linear limb darkening coefficients from 0.3 to 0.4 and integrated over the limb darkening coefficients to obtain a marginalized likelihood function. We were left with a likelihood function that depends only on the parameter of interest ($\alpha$). We plot the likelihood functions for different values of $\mu$ and the marginalized likelihood function in Figure \ref{f:limbDark}. 

\section{Results} 
\label{results} 

\begin{figure}[ht]
\begin{center}
\includegraphics[width=\linewidth]{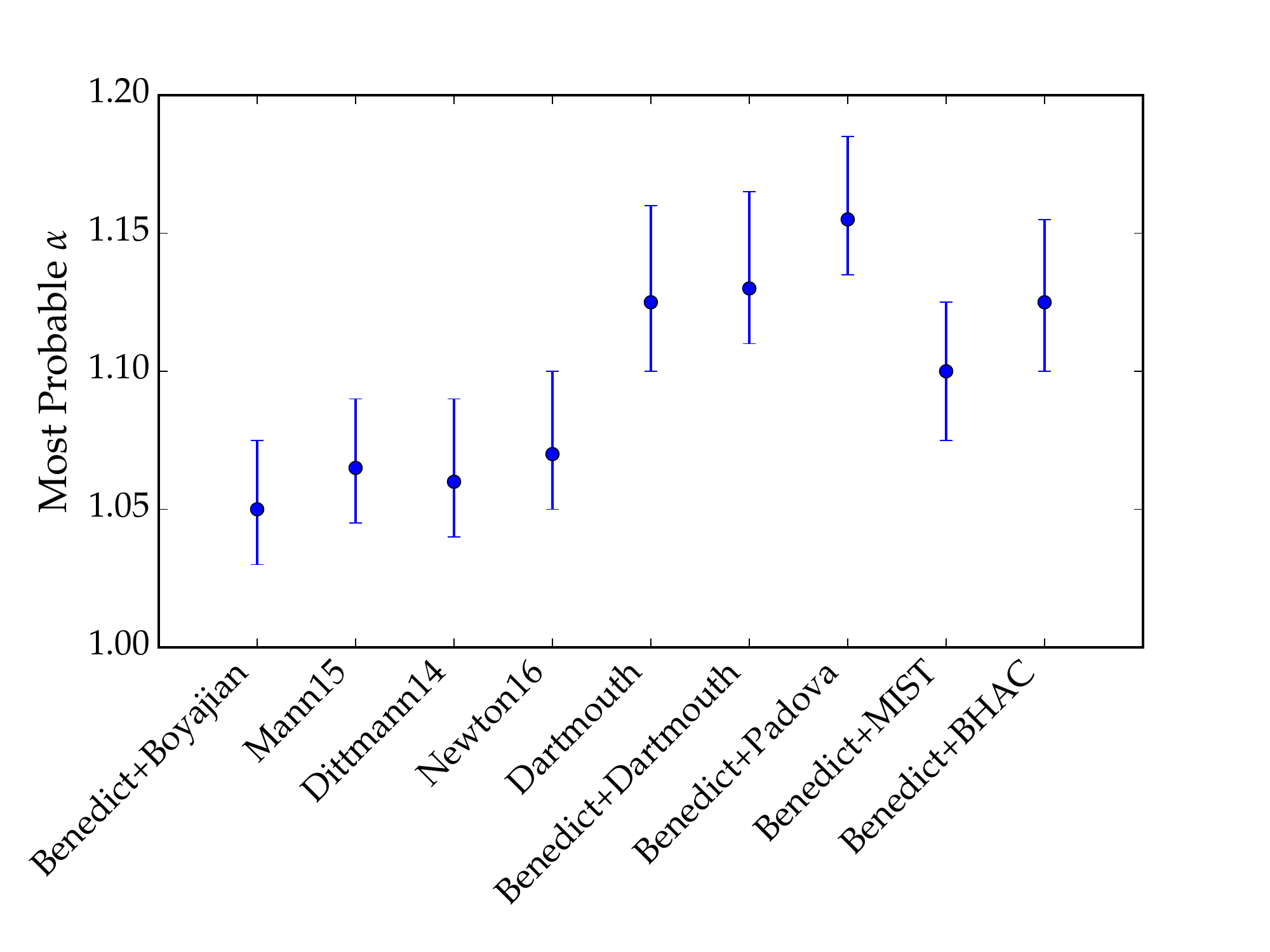}
\caption{\small
Results from the marginalized likelihood PDFs for the different radius estimates in Table \ref{t:radiiMethods}. The central blue point for each method denotes where the peak of the likelihood function falls. The error bars are one sigma error bars and show where 68\% of the combined probability lies. All of the methods that use empirical relations instead of stellar evolutionary models show significantly lower levels of discrepancy between the data and the predicted radius values. }
\label{f:radiusMethodCompare}
\end{center}
\end{figure}

We performed the same steps of constructing a likelihood function but used the other published radius values and relation instead of those published in \citet{dittmann14}. 
The results of the Bayesian analysis for each method are shown in Figure \ref{f:radiusMethodCompare}. The method that shows the least amount of discrepancy between the observed data and results is Benedict+Boyajian, which combines the most recent empirically derived mass and radius relations. All of the empirical relations show better agreement between the observed data and radius predictions than the radius predictions that utilize stellar evolutionary models. 

To determine the statistical significance of whether an inflated model is preferred, we employed both the odds ratio and the Bayesian information criterion (BIC). The odds ratio tests the relative probabilities of two theories and takes into account both likelihoods and priors. This serves to penalize theories that are more complex and explore more parameter space, however it can be a problem if the prior is not well defined, as different priors can significantly change the odds ratio. As stated previously we will use a uniform prior since we do not have much specific prior information on inflation. In this case, the odds ratio is equal to the Bayes factor ($B_{10}$). The BIC on the other hand is an \textit{approximation} of the log of the Bayes factor, but does not require a prior. It still penalizes complex theories, however, by taking into account the number of free parameters present. 

We used the equation for the Bayes factor derived in Eq. 3.24 of \citet{gregory}: 

\begin{equation}
B_{10} \approx \frac{L(\hat{\alpha})\delta \alpha}{L(\alpha_{0})\Delta \alpha}
\end{equation}

\noindent where $\alpha$ is the free parameter (here inflation percent), $L(\hat{\alpha}$) is the likelihood at the maximum inflation, $L(\alpha_{0}$) is the likelihood at 0\% inflation, $\delta \alpha$ is the RMS about the maximum inflation of the likelihood function, and $\Delta \alpha$ is the width of the uniform prior. We tried two different priors. For prior 1 we chose the inflation range from $0-15\%$ radius inflation because these are the results often quoted from EBs \citep[e.g., ][]{torres02}. For the second prior we use the entire explored range of parameter space (from $-10\% to 25\%$).  

To calculate the BIC we use the Schwarz criterion as stated in \citet{kass95}: 

\begin{equation} 
BIC = -2\times ln(L(\alpha)) + k \times ln(N)
\end{equation}
where k is the number of free parameters (one for the model with inflation and zero for the model with a fixed inflation of 0\%), and N is the sample size. We can then calculate the BIC for both $L(\hat{\alpha}$) and $L(\alpha_{0}$) and subtract them to get $\Delta$BIC. As stated in \citet{kass95} $\Delta$BIC is approximately equal to two times $log_{e}$ of the Bayes Factor. \citet{kass95} also provide a detailed analysis of how both the Bayes factor and the BIC translate to statements of statistical significance.  

Finally, to allow for easier interpretation of the results, we translate our Bayes Factors into frequentist $p$-values using the equation $B_{ij} = -(e\ p\ ln(p))^{-1}$, where $p$ is the $p$-value and $p < e$ \citep{sellke01}. The results are summarized in Table \ref{t:statsResults}. We find that all three predictions that involve stellar evolutionary models show `Strong' to `Very Strong' evidence that the observed M dwarf stars are larger than model radius estimates. The radii reported in both \citet{newton16} and \citet{dittmann14} show 2- to 3-sigma levels of discrepancy between the quoted radii and the measured radii (where the measured radii are on average $6-7\%$ larger than reported radii). However, when we use the newest empirical relations from \citet{benedict16} and \citet{boyajian12} we find that both of the odds ratios and the BIC cannot rule out the null hypothesis, that there is no inflation. Even though the maximum likelihood occurs for radii 5\% larger than the relations predict, the increase in total probability is not enough to overcome the penalty imposed by adding a free parameter. %Therefore, we can conclude that radii of rapidly rotating and magnetically active stars can be estimated to better than 5\% by using the \citet{benedict16} and \citet{boyajian12} empirical relations. 

\begin{deluxetable*}{c c c c c c c c}
\tablecolumns{8}
\centering
\tablewidth{0pt}
\tablecaption{Significance of Radius Inflation \label{t:statsResults}}%title of the table %title of the table
\tabletypesize{\footnotesize}
\tablehead{\colhead{Method Name} & \colhead{Radius Under-} & \colhead{Odds Ratio} & \colhead{P-value} & \colhead{Odds Ratio} & \colhead{P-value} & \colhead{$\Delta$BIC} & \colhead{Statement of}
\\ 
\colhead{} & \colhead{prediction (\%)} & \colhead{(Prior 1)} & \colhead{(Prior 1)} & \colhead{(Prior 2)}& \colhead{(Prior 2)} & \colhead{} & \colhead{Significance}}

\startdata
Benedict+Boyajian & $5 \substack{+2.5 \\ -2 }$ & 0.991 & - & 0.431 & - & -0.424 & No evidence of inflation \\
[0.2cm]
Mann15 & $6.5 \substack{+2.5 \\ -2 }$ & 3.33 & 0.032 & 1.45 & 0.12 & 0.788 & Positive evidence of inflation \\ [0.2cm]
Dittmann14 & $6 \substack{+3 \\ -2 }$ & 2.966 & 0.038 & 1.290 & 0.15 & 0.567 & Positive evidence of inflation\\
[0.2cm]
Newton16 & $7 \substack{+3 \\ -2 }$ & 8.882 & 0.009 & 3.862 & 0.026 & 1.663 & Positive evidence of inflation \\
[0.2cm]
Dartmouth & $12.5 \substack{+3.5 \\ -2.5 }$ & $2.9\times10^3$ & $1.11\times10^{-5}$ & $1.27\times10^3$ & $2.76\times10^{-5}$ & 7.276 &  Strong evidence of inflation\\
[0.2cm]
Benedict+Dartmouth & $13 \substack{+3 \\ -2.5 }$ & $8.37\times10^4$ & $2.92\times10^{-7}$ & $3.64\times10^4$ & $7.14\times10^{-7}$ & 10.72 & Very Strong evidence of inflation \\ 
[0.2cm]
Benedict+Padova & $16.5 \substack{+3 \\ -2 }$ &$1.48\times10^7$ & $1.21\times10^{-9}$ & $6.45\times10^6$ & $2.9\times10^{-9}$ & 16.0 & Very Strong evidence of inflation \\ 
[0.2cm]
Benedict+MIST & $10 \substack{+3 \\ -2 }$ &$240.8$ & $1.76\times10^{-4}$ & $104.7$ & $4.57\times10^{-4}$ & 4.96 & Strong evidence of inflation \\ 
[0.2cm]
Benedict+BHAC & $12.5 \substack{+3 \\ -2 }$ &$1.97\times10^4$ & $1.38\times10^{-6}$ & $8.57\times10^3$ & $3.41\times10^{-6}$ & 9.27 & Strong evidence of inflation \\ 
\enddata

\end{deluxetable*}

\section{Potential Biases}
\label{biases}

To ensure that these results were accurate and that there was not a bias in the sample, or a bias that occurred when combining a rotational period with rotational broadening, we explored all of the possibilities we imagined where this could occur.

\subsection{Differential Rotation}

\begin{figure}[ht]
\begin{center}
\includegraphics[width=\linewidth]{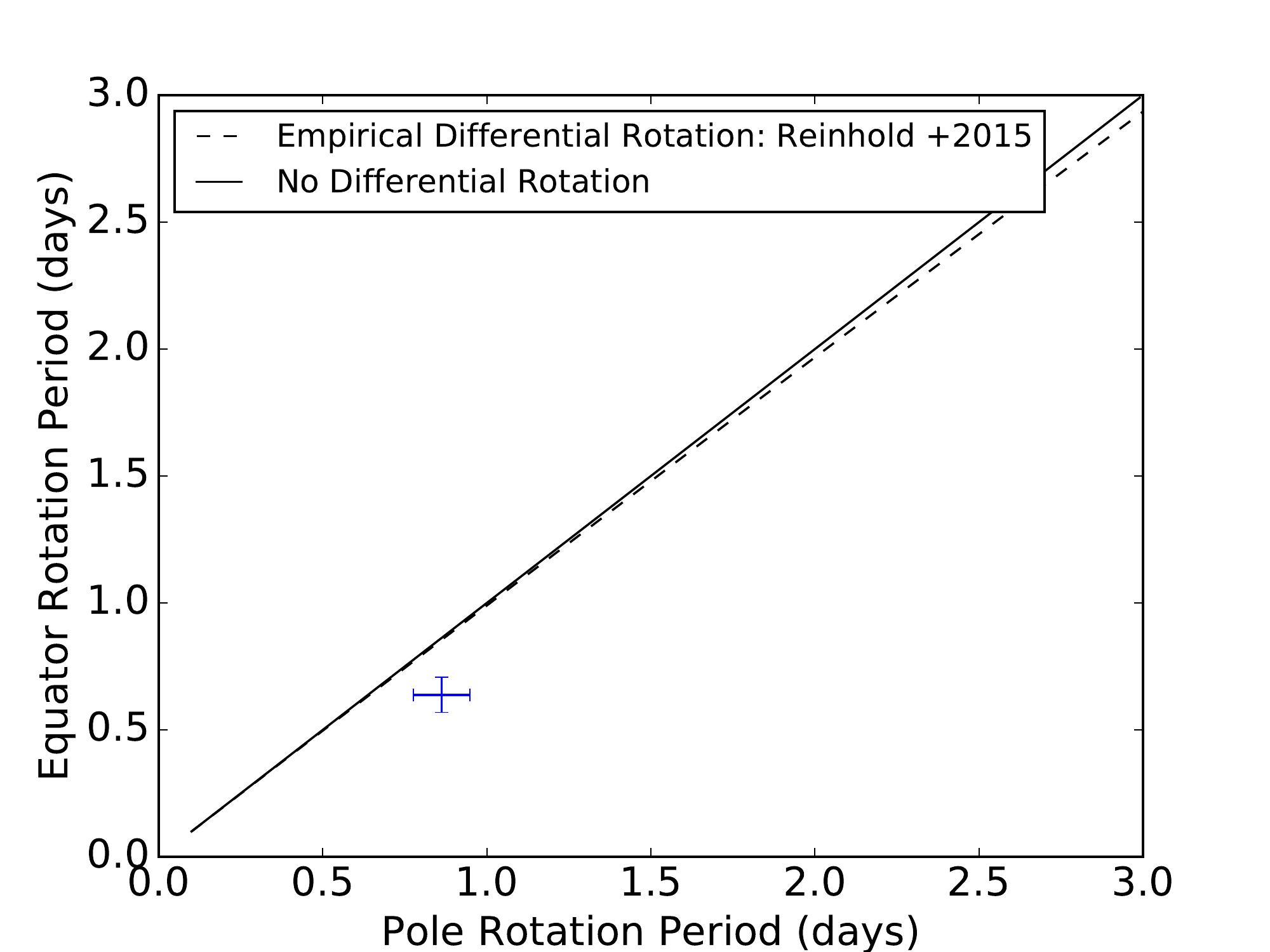}
\caption{\small
The polar rotation period (minimum rotation period) determined through spot modulation and reported in \citet{newton16} versus the equatorial rotation period (maximum rotation period) determined through our $v \sin i$ measurements. The solid line, where both periods are the same, is the expected result for no differential rotation. The dotted line shows the relation from \citet{reinhold15}. The empirical relation from \citet{kuker11} has even smaller deviations from the line showing no differential rotation, therefore we do not display it on our plot. A target with a large deviation between the rotational and $v \sin i$ period is shown in blue, with observed errors. To calculate an equivalent period from the measured $v \sin i$ we assume $\sin i = 90\arcdeg$, which gives the largest equivalent period (i.e., minimizes the difference between the two periods). We can conclude that differential rotation cannot account for the observed discrepancy. }
\label{f:diffRot}
\end{center}
\end{figure}

Because spots are primarily located at high latitudes on M dwarf stars \citep{barnes15}, and $v \sin i$ measurements are primarily sensitive to equatorial rotation, any discrepancies between the measured $v \sin i$ values and spot modulation periods could be due to differential rotation. However, both observations and models of differential rotation on low-mass, rapidly rotating stars yield extremely small shear values and cannot account for the observed discrepancies that we found. Using $Kepler$ data of more than 10,000 stars, \citet{reinhold15} showed that there is a relationship between the horizontal rotation shear and the rotation period, where stars with faster rotation periods exhibited smaller shears. \citet{reinhold15} also found that stars categorized as having the most stable rotation period (deviations less than 0.001 days) all had periods of less than 10 days, and the distribution peaked at periods less than 1 day. This same result has been shown previously with smaller data sets \citep[e.g., ][]{hall91, donahue96}. A relationship between the differential rotation and the effective temperature was observed by \citet{barnes05}, where stars with cooler effective temperatures were found to have less differential rotation. Models of differential rotation provide further evidence to these observational findings and show that the shear decreases with decreasing rotation period, and with decreasing mass for a rotation period that is held constant \citep{kuker11}.  Therefore our sample of low-mass rapidly rotating stars should have very little, if any differential rotation since we are probing the parameter space least affected by rotational shears. In figure \ref{f:diffRot} we plot the relation from \citet{reinhold15} to show that differential rotation cannot account for the larger rotational broadening values compared to rotational periods. 

\subsection{Isochrone Age and Metallicity} 
\label{DartBias}

Since the stars in our sample are rapid rotators and are magnetically active, they are also likely young. There is evidence that M dwarf stars do not follow an exact Skumanich-like relation between the rotation period and age \citep{skumanich72}, but instead a rotation period dichotemy exists \citep{newton16}. Fully convective M dwarf stars can continue to be magnetically active and retain rotation periods of less than 10 days up until $5-7$Gyrs \citet{west08}, then it appears that they shed angular momentum and rapidly migrate to periods greater than $\sim$30 days \citep{newton16}. This makes precise gyrochonology very difficult for these stars, however it is well established that rapid rotators are on average younger than slow rotating M dwarfs \citep{west08, west15}.  We therefore do not explore using isochrones with ages larger than 5 Gyrs. 

We explore many scenarios with isochrones of younger ages. Using a 1 Gyr isochrone from the Dartmouth models, we find almost the same likelihood function, however with a 0.5\% increase to even higher levels of inflation.
Both the MIST and BHAC models offer isochrone grids down to ages of a few million years. We find changes of less than 1\% in the most likely inflation at an age of 500 Myrs for both sets of models. At 250 Myrs, the MIST models show an $\alpha$ of 4.5\%, which is no longer statistically significant. Performing the same analysis for the BHAC models, we find that at 200 Myrs we still measure an $\alpha$ value of 9\%, and it is not until 120 Myrs that we no longer measure a statistically significant value of $\alpha$. We assert that it is highly unlikely that all of the stars in our sample are this young since none of the stars are associated with star clusters or moving groups and parallax measurements from \citet{dittmann14} indicate that the stars are located on the main-sequence. Further evidence that age is not the sole contributing factor of the observed inflation is given by comparison with rotation periods observed in young clusters such as Pleiades and NGC 2516. For mid-to-late M dwarfs neither of these young ($\sim120-150$ Myrs) clusters are observed to contain stars with rotation periods longer than about 1.5 to 2 days \citep{scholz11,rebull16,rebull16b}, however many of the stars in our sample that have the largest observed mismatch between the rotation period and the rotational broadening have rotation periods in the 1-5 day period regime and are therefore probably older than 150 Myrs. It is possible that some of the stars have ages of $200-300$ Myrs since they would be almost indistinguishable from main-sequence stars and some of the measured inflation could be due to age, however this would not explain the similarity between the $R \sin i$ distribution and radii from interferometry, which are measured on older, slowly rotating stars. Therefore, we assert that the majority of the observed inflation is not due to age. 

In a study of the metallicity of the MEarth sample \citep{newton14}, the average metallicity of the rapidly rotating stars is $0.14 \pm 0.1 dex$. Therefore in our analysis we assumed a metallicity of 0.14 dex when comparing to isochrones. We find that by using a solar metallicity isochrone the average inflation can change by $1-1.5\%$. Since this change in metallicity is more than one standard deviation and it can only account for a small amount of the observed inflation, this leads us to conclude that metallicity alone cannot be responsible for the inflation observed in our stellar sample. We note that the metallicities were measured using methods that may in fact be probing the carbon-to-oxygen ratios of the stars, and not the metallicities directly \citep{veyette16, veyette17}.

\subsection{Microturbulence} 

Microturbulence is another broadening mechanism in the spectra of stars, and some of the broadening we measure could be due to microturbulence and not rotational broadening. If microturbulence affected the spectra of our slowly rotating templates and the rapid rotators to the same degree this would not be a problem, however microturbulence could potentially affect the spectra of the young rapid rotators to a greater degree.
We performed a simple order of magnitude test to determine how much microturblence would be required to relieve the 5-6\% discrepancy between empirical relations and our $R \sin i$ measurements. For a simple order of magnitude estimation, we can assume that microturbulence and rotational broadening add in quadrature. We can then estimate that the total broadening ($v_{tot}$) is related to the broadening from microturblence and rotation as follows: $v_{tot} = \sqrt{v_{rot}^2 + v_{micro}^2}$. 
We find that in order to negate a 5\% offset between data and empirical relations or models, microturblence needs to contribute 4 km/s of broadening. This does not seem likely that the entire offset between empirical relations and our measured $R \sin i$ values is due to microturbulence because it is estimated that microturbulence contributes 1-2 km/s of broadening to low-mass stars \citep{reid05}. However, this 1-2 km/s of broadening would account for about $0.5-1.5\%$ of the discrepancy between the radius prediction methods and our data, and the true values of $\alpha$ for each method (see Figure \ref{f:radiusMethodCompare}) could be about 0.005-0.015 smaller.

\section{Discussion and Conclusions}
\label{conclusions}

\begin{figure}[ht]
\begin{center}
\includegraphics[width=\linewidth]{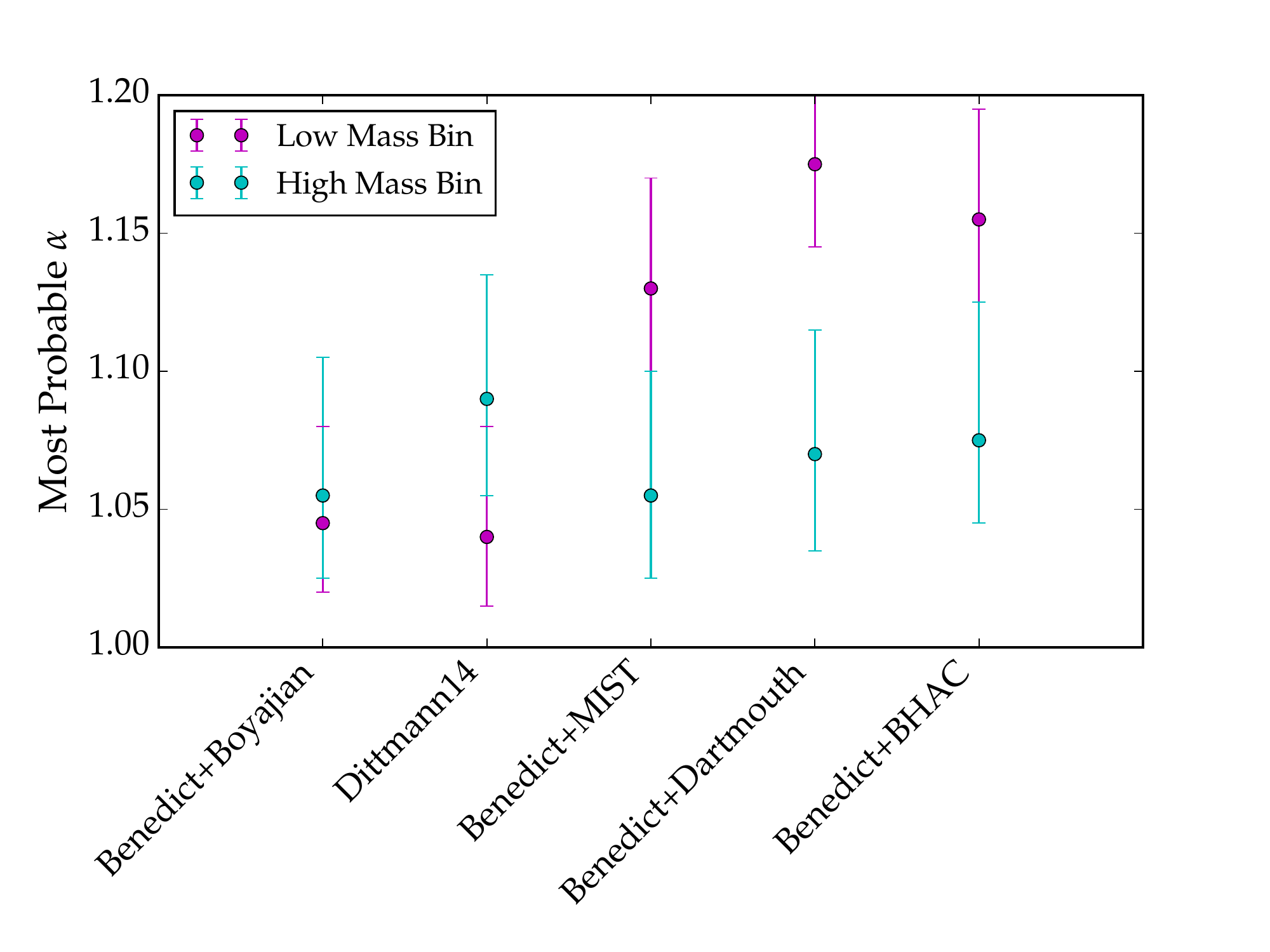}
\caption{\small
Same as Figure \ref{f:radiusMethodCompare}, but with the stellar sample split into two similarly sized mass bins. The lower mass bin contains stars with $0.08 M_\odot< M <0.18 M_\odot$, while the higher mass been contains stars with $0.18 M_\odot< M <0.4 M_\odot$. We find that the lower mass stars are significantly more inflated than the higher mass stars when compared to models. The lower mass stars are inflated by $13 \substack{+4 \\ -3}\%$ compared to the MIST models and $17.5 \substack{+3.5 \\ -3}\%$ compared to the Dartmouth models, and $15.5\substack{+4 \\ -3}\%$ compared to the BHAC models. The higher mass stars are only inflated by $5.5 \substack{+4.5 \\ -3}\%$ compared to MIST models, $7 \substack{+4.5 \\ -3.5}\%$ compared to the Dartmouth models, and $7.5\substack{+5 \\ -3}\%$ compared to the BHAC models. The empirical relations do not seem to show the same trend that the lower mass stars are more inflated than the higher mass stars and for both empirical relations the points are within one standard deviation of each other.}
\label{f:lowHighLikelihood}
\end{center}
\end{figure}

We find that stellar evolutionary models under-predict radii of our sample of low-mass stars by between 10 and 16.5\% depending on the model, and that including radius inflation is strongly favored over model predictions without radius inflation. This is higher than the average inflation seen in EB systems \citep[$\sim5\%$ from a literature compilation in][]{han17}, so we decided to test if this inflation was consistent over the whole mass range. We split the data into two mass bins of roughly equal numbers of targets, one with stars that had $0.08 M_\odot< M <0.18 M_\odot$ and the second that had $0.18 M_\odot< M <0.4 M_\odot$. We then computed separate likelihood functions for each of these; the results are shown in Figure \ref{f:lowHighLikelihood}. We find that the higher mass bin has an average radius inflation of $5-7 \substack{+4.5 \\ -3.5}\%$, which is consistent with results from EBs. In the lower mass bin we find the average inflation is $13-17.5 \substack{+4 \\ -3}\%$. In this low mass range there are very few known EBs and only two stars with long baseline optical interferometry measurements with which to calibrate models. 

The inflation seen here for the higher mass bin is consistent to that observed in partially convective EBs, so we can conclude that radius inflation is not a symptom of binarity (or how parameters are extracted from EBs). We also find that there is no significant change in the amount of inflation compared to models across the fully convective boundary, and that our higher mass bin shows similar levels of inflation as partially convective stars. However, for stars at the very end of the main sequence, stellar evolutionary models severely underestimate stellar radii. While this could be an issue of age (the lowest mass stars have not evolved onto the main sequence yet and are still contracting), it is also possible that models of the lowest-mass stars are inaccurate. More work is needed to validate this result and to test why stellar evolutionary models underestimate the radii of the lowest mass stars by 15-20\%.

Since partially convective and fully convective stars are inflated by similar amounts, we can provide constraints to modeling efforts. It is still disputed in the literature as to whether strong magnetic fields can inhibit convection and inflate radii in fully convective stars to the $\sim$10\% seen here and in EBs. 
\citet{macdonald17} state that they can produce radius inflation at the $\sim10\%$ level by modeling the stabilization of convection with magnetic fields on the order or 10kG, while \citet{feiden14} argue that using a similar method, they require unreasonably large magnetic fields to inflate the radii by even 5\%. 
Our data are consistent with the results from \citet{macdonald17}, but in the scenario put forth by \citet{feiden14}, magnetic spots would be required to produce the observed inflation in fully convective stars. More exploration of spot modeling would increase our understanding of the problem and help distinguish between the two modeling frameworks.  

Radii reported in \citet{newton16} and \citet{dittmann14}, and radii calculated using the relations in \citet{mann15} seem to under-predict our sample by 6-7\%, but only with a moderate level of statistical significance ($2 - 3 \sigma$). When we use the most recent empirical $M_{K} -$Mass relation \citep{benedict16} and Mass$-$Radius relation \citep{boyajian12}, we find no statistically significant evidence that a model with inflation describes the data better than a model without inflation. The Mass$-$Radius relation used to determine these radii was calibrated using slowly rotating stars.
Using this relationship on our rapidly rotating sample returns statistically consistent results, leading us to conclude that if rotation inflates the radii of fully convective rapidly rotating stars, it seems to be less than $5\substack{+2.5 \\ -2 }\%$.

Further evidence that rotation does not significantly effect the radii is given by the fact that slowly and rapidly rotating stars seems to be inflated by similar amounts compared to models. We calculated updated mass estimates for Proxima Centauri and Barnard's Star using $K$-band magnitudes and distances reported in \citet{boyajian12}, and applying the $M_{K}-Mass$ relation from \citet{benedict16}. We then used a relation from the Dartmouth code for solar metallicity and ages of 5 Gyrs and 10 Gyrs for Proxima Centauri and Barnard's Star, and found models underestimated the radii for both stars by 3-4\% compared to the optical interferometry radius measurements from \citet{boyajian12}. Further evidence of slowly rotating mid-to-late M dwarf stars with inflated radii was noted by \citet{irwin11}, who measured the radii of a long period (41 days) EB and found the component radii to be inflated by 4\%. Our bin of higher mass stars is inflated by $5-7 \substack{+4.5 \\ -3.5}\%$, which is consistent with 3-5\% radius inflation of slowly rotating stars.

We conclude that the \citet{benedict16} and \citet{boyajian12} relations are accurate (to an uncertainty of $\sim5\%$) for rapidly rotating, magnetically active, fully-convective M dwarf stars. These relations have not been thoroughly tested at the very low-mass end of the main sequence. \citet{boyajian12} explicitly warn that their relations may not be accurate for spectral types later than M4. We can therefore provide evidence that the relations hold to within uncertainties of $\sim5\%$ at the end of the main sequence ($M \sim 0.08M_\odot$) for the most rapidly rotating and magnetically active stars.

Finally, we note that with many of the upcoming missions, this type of analysis can be performed on larger samples of stars that cover a wider parameter space in the future. \textit{GAIA} parallaxes as well as photometric rotation periods from surveys such as \textit{TESS} and the Large Synoptic Survey Telescope (LSST) will be available within the next few years and will increase the number or potential targets by orders of magnitude. This method is especially promising for brown dwarf radii since there are almost no known brown dwarf EBs and brown dwarfs are too dim for long baseline optical interferometry. A sample of brown dwarfs with known rotation periods and parallaxes will allow us to observationally constrain models and radius estimates. 

\acknowledgements
The authors would like to thank the referee for the thoughtful report, which greatly improved the manuscript.
The authors would also like to thank Lisa Prato and Larissa Nofi for IGRINS training, and Heidi Larson, Jason Sanborn, and Andrew Hayslip for operating the DCT during our observations. We would also like to thank Jen Winters, Jonathan Irwin, Paul Dalba, Mark Veyette, Eunkyu Han, and Andrew Vanderburg for useful discussions and helpful comments on this work. 
Some of this work was supported by the NASA Exoplanet Research Program (XRP) under Grant No. NNX15AG08G issued through the Science Mission Directorate.

These results made use of the Lowell Observatory's Discovery Channel Telescope, supported by Discovery Communications, Inc., Boston University, the University of Maryland, the University of Toledo and Northern Arizona University; the Immersion Grating Infrared Spectrograph (IGRINS) that was developed under a collaboration between the University of Texas at Austin and the Korea Astronomy and Space Science Institute (KASI) with the financial support of the US National Science Foundation under grant AST-1229522, of the University of Texas at Austin, and of the Korean GMT Project of KASI; data taken at The McDonald Observatory of The University of Texas at Austin; and data products from the Two
Micron All Sky Survey, which is a joint project of the
University of Massachusetts and the Infrared Processing and
Analysis Center/California Institute of Technology, funded by
NASA and the NSF.

\facilities{DCT (IGRINS), IRTF (iSHELL), Smith (IGRINS)}

\software{PyAstronomy, Spextool \citep{cushing04}}

\begin{longrotatetable}
\renewcommand\thefootnote{\alph{footnote}} 
\begin{deluxetable}{c c c c c c c c c c l}
\tablecolumns{5}
\centering
\tablewidth{0pt}
\tablecaption{Observed Targets \label{t:obsData}} %title of the table
\tablehead{\colhead{2MASS name} & \colhead{Observation} & \colhead{Telescope} & \colhead{Instrument} & \colhead{Rotation} & \colhead{$v \sin i$} & \colhead{$\sigma_{v \sin i}$} & \colhead{Previous} & \colhead{Previous} & \colhead{Reference}
\\ 
\colhead{} & \colhead{Date (UT)} & \colhead{} & \colhead{} & \colhead{Period (d)\footnote[1]{\citet{newton16}}}& \colhead{(km/s)} & \colhead{(km/s)} & \colhead{$v \sin i$} & \colhead{$\sigma_{v \sin i}$} & \colhead{}}
\startdata
J00243478+3002295&24 Sept 2017 &DCT&IGRINS&1.077&13.0&0.5 & 12.2 & 0.8 & F17\footnote[2]{\citet{fouque17}} \\ 
J00304867+7742338&24 Sept 2017&DCT&IGRINS&0.137&30.5&0.9 & & & \\ 
J00544803+2731035&2 Aug 2017&IRTF&iSHELL&1.697&9.0&0.7 & & & \\ 
J01015952+5410577&25 Sept 2017&DCT&IGRINS&0.278&31.9&0.9 & 30.6 & 3.1 & R17\footnote[3]{\citet{reiners17}}\\ 
J01533076+0147559&10 Nov 2017&DCT&IGRINS&0.199&34.3&3.3 & & & \\ 
J01534955+4427284&26 Sept 2017&DCT&IGRINS&0.216&47.6&2.9 & & & \\ 
J01564570+3033288&2 Aug 2017&IRTF&iSHELL&1.581&6.7&0.8 & & & \\
J01584517+4049445&24 Sept 2017&DCT&IGRINS&0.486&21.8&0.5 & & & \\ 
J02032864+2134168&11 Nov 2017&DCT&IGRINS&0.32&27.7&1.5 & & & \\ 
J02071032+6417114&1 Sept 2017&IRTF&iSHELL&1.177&11.3&0.9 & 11.4 & 1.0 &F17 \\ 
J02170993+3526330&25 Sept 2017&DCT&IGRINS&0.276&23.5&0.7 & 28.2 & 0.7 & J09\footnote[4]{\citet{jenkins09}}\\ 
J02204625+0258375&24 Sept 2017&DCT&IGRINS&0.503&20.7&0.4 & 23.3 & 0.7 & J09 \\ 
J02351494+0247534&26 Sept 2017&DCT&IGRINS&0.472&7.9&0.7 & & & \\ 
J02364412+2240265&25 Sept 2017&DCT&IGRINS&0.37&12.7&1.4 & 11.2&1.4 &F17 \\ 
J02514973+2929131&26 Sept 2017&DCT&IGRINS&0.895&19.2&1.7 & & & \\ 
J03205965+1854233&24 Sept 2017&DCT&IGRINS&0.614&8.4&1.1 & 8.0 & - & R02\footnote[5]{\citet{reid02}} \\ 
J03284958+2629122&6 Nov 2017&IRTF&iSHELL&3.235&2.8&0.4 & & & \\ 
J03304890+5413551&14 Nov 2017&DCT&IGRINS&0.117&47.7&2.1 & & & \\ 
J03360868+3118398&24 Sept 2017&DCT&IGRINS&0.856&15.7&0.4 & & & \\ 
J03425325+2326495&26 Sept 2017&DCT&IGRINS&0.834&8.3&0.7 & 12.7& 0.5 & D13\footnote[6]{\citet{deshpande13}}\\ 
J03571999+4107426&12 Nov 2017&DCT&IGRINS&0.567&6.5&0.8 & & & \\ 
J04121693+6443560&6 Nov 2017&IRTF&iSHELL&1.594&7.2&0.2 & & & \\ 
J04140201+8215360&10 Nov 2017&DCT&IGRINS&0.277&17.0&0.8 & & & \\ 
J04171852+0849220&25 Sept 2017&DCT&IGRINS&0.185&37.3&1.1 &&&\\ 
J04201254+8454062&12 Nov 2017&DCT&IGRINS&0.695&15.2&0.7&&&\\ 
J04302527+3951000&12 Nov 2017&DCT&IGRINS&0.718&14.2&0.5&13.6&0.8&F17\\ 
J04333393+2044461&11 Nov 2017&DCT&IGRINS&0.335&27.1&2.0&&&\\ 
J04434430+1505565&14 Nov 2017&DCT&IGRINS&0.419&22.3&0.5&&&\\ 
J04490464+5138412&12 Nov 2017&DCT&IGRINS&0.724&9.7&0.7&&&\\ 
J05041476+1103238&11 Nov 2017&DCT&IGRINS&0.842&10.8&1.4&&&\\ 
J05062489+5247187&10 Nov 2017&DCT&IGRINS&0.648&14.5&1.1&&&\\ 
J05405390+0854183&10 Nov 2017&DCT&IGRINS&0.332&15.5&0.6&&&\\ 
J05595569+5834155&11 Nov 2017&DCT&IGRINS&0.951&9.2&1.7&&&\\ 
J06000351+0242236&6 Nov 2017&IRTF&iSHELL&1.809&5.7&0.4&5.8, 5.9&0.3, 1.4&D15\footnote[7]{\citet{davison15}}, F17\\ 
J06052936+6049231&10 Nov 2017&DCT&IGRINS&0.31&29.7&2.6&&&\\ 
J06073185+4712266&28 Jan 2017&DCT&IGRINS&0.862&20.8&0.8&&&\\ 
J06235123+4540050&6 Nov 2017&IRTF&iSHELL&2.515&6.7&0.6&&&\\ 
J06481555+0326243&12 Nov 2017&DCT&IGRINS&0.458&9.2&0.7&&&\\ 
J07454039+4931488&11 Nov 2017&DCT&IGRINS&0.253&18.5&1.8&&&\\ 
J07464203+5726534&30 Jan 2017&DCT&IGRINS&0.82&17.6&0.8&&&\\ 
J07555396+8323049&11 May 2017&Harlan J. Smith&IGRINS&1.107&13.4&1.2&&&\\ 
J08012112+5624042&11 Nov 2017&DCT&IGRINS&0.117&66.0&0.5&&&\\ 
J08055713+0417035&10 Nov 2017&DCT&IGRINS&0.176&29.7&1.5&&&\\ 
J08212804+5220587&12 Nov 2017&DCT&IGRINS&0.472&12.7&0.4&&&\\ 
J08294949+2646348&28 Jan 2017&DCT&IGRINS&0.459&9.6&0.8&8.1, 10.5, 11.4&1.1, 1.5, 0.7& D98\footnote[8]{\citet{delfosse98}}, R17, F17\\ 
J08505062+5253462&6 Nov 2017&IRTF&iSHELL&1.754&9.6&1.1&13.1&0.7&J09\\ 
J08593592+5343505&29 Jan 2017&DCT&IGRINS&0.581&26.9&0.8&&&\\ 
J09002359+2150054&30 Jan 2017&DCT&IGRINS&0.439&15.5&0.4&20.0, 14.3, 15.0&0.6, 1.5, 1.0&J09, R17, F17\\ 
J09245082+3041373&10 Nov 2017&DCT&IGRINS&0.373&44.9&3.1&&&\\ 
J09535523+2056460&30 Jan 2017&DCT&IGRINS&0.615&10.1&0.6&16.5&0.4&J09\\ 
J09585650+0558000&29 Jan 2017&DCT&IGRINS&0.453&22.3&0.4&&&\\ 
J09591880+4350256&30 Jan 2017&DCT&IGRINS&0.755&22.5&0.7&&&\\ 
J10011109+8109226&11 May 2017&Harlan J. Smith&IGRINS&0.302&21.3&3.1&&&\\ 
J10024936+4827333&29 Jan 2017&DCT&IGRINS&0.268&18.4&0.4&&&\\ 
J10030191+3433197&30 Jan 2017&DCT&IGRINS&0.859&11.9&0.2&&&\\ 
J10204406+0814234&10 Nov 2017&DCT&IGRINS&1.087&17.1&1.0&&&\\ 
J10252645+0512391&12 Nov 2017&DCT&IGRINS&0.102&59.9&2.1&&&\\ 
J10521423+0555098&30 Jan 2017&DCT&IGRINS&0.692&13.6&0.4&19.1&0.2&J09\\ 
J11005043+1204108&30 Jan 2017&DCT&IGRINS&0.298&28.6&1.3&26.5&0.8&D13\\ 
J11224274+3755484&8 May 2017&Harlan J. Smith&IGRINS&0.358&13.3&0.8&&&\\ 
J11432359+2518137&10 Nov 2017&DCT&IGRINS&1.326&13.5&0.9&13.7&0.9&F17\\ 
J11483548+0741403&28 Jan 2017&DCT&IGRINS&0.708&14.0&0.5&&&\\ 
J12041256+0514128&29 Jan 2017&DCT&IGRINS&0.154&23.0&0.5&&&\\ 
J12185939+1107338&30 Jan 2017&DCT&IGRINS&0.491&16.3&0.4&9.2, 15.6&1.9, 0.8&D98, F17\\ 
J12265737+2700536&28 Jan 2017&DCT&IGRINS&0.733&4.0&0.7&13.5&0.6&D13\\ 
J13003350+0541081&29 Jan 2017&DCT&IGRINS&0.6&16.9&0.5&16.8, 15.6&2.1, 0.8&D98, F17\\ 
J13093495+2859065&28 Jan 2017&DCT&IGRINS&0.215&48.6&1.1&51.3&1.5&F17\\ 
J13533877+7737083&- & - & - & 1.231 & - & - & 8.9 & 1.5 & R17 \\
J14224340+1624464&29 Jan 2017&DCT&IGRINS&0.889&8.0&0.5&&&\\ 
J14311348+7526423&30 Jan 2017&DCT&IGRINS&0.631&14.3&0.4&&&\\ 
J15163731+5355457&30 Jan 2017&DCT&IGRINS&0.525&19.2&0.4&&&\\ 
J15164073+3910486&30 Jan 2017&DCT&IGRINS&0.581&16.3&0.4&&&\\ 
J16400599+0042188&- & - & - & 0.311 & - & - & 31.0 & 0.8 & F17 \\
J16402068+6736046&- & - & - & 0.378 & - & - & 10.8 & 0.7 & F17 \\
J18021660+6415445&11 May 2017&Harlan J. Smith&IGRINS&0.28&10.3&0.9&11.3, 13.2&1.5, 1.2&R17, F17\\ 
J18315610+7730367&- & - & - & 0.861 & - & - & 15.8 & 0.7 & F17 \\
J18481752+0741210&- & - & - & 2.756 & - & - & 2.4 & 1.5 & R17 \\
J19510930+4628598&May 2017&Harlan J. Smithd&IGRINS&0.593&22.9&0.5&22.1&0.9&F17\\ 
J20045709+0321076&17 Oct 2016&DCT&IGRINS&0.788&12.0&0.5&&&\\ 
J22482247+1232105&17 Oct 2016&DCT&IGRINS&0.633&7.7&1.1&&&\\ 
J22502051+5136265&24 Sept 2017&DCT&IGRINS&0.883&11.0&0.5&&&\\ 
J22541111+2527562&18 Oct 2016&DCT&IGRINS&0.356&8.7&0.5&&&\\ 
J23025250+4338157&18 Oct 2016&DCT&IGRINS&0.348&29.0&1.5&&&\\ 
J23270216+2710367&25 Sept 2017&DCT&IGRINS&0.922&13.0&0.5&&&\\ 
J23310587+0842314\footnote[9]{Spectroscopic Binary}&25 Sept 2017&DCT&IGRINS&1.647&9.1&0.8&&&\\ 
J23383392+0624518&19 Oct 2016&DCT&IGRINS&0.251&31.7&1.1&&&\\
J23512227+2344207&- & - & - & 3.211 & - & - & 5.2 & 0.9 & F17 \\
J23545147+3831363&- & - & - & 4.755 & - & - & 3.6 & 1.5 & R17 \\
\enddata

\end{deluxetable}
\end{longrotatetable}

\end{document}